\documentclass[manuscript,screen,nonacm]{acmart}

\AtBeginDocument{%
  }

\acmISBN{978-1-4503-XXXX-X/2018/06}

\usepackage{booktabs}
\usepackage{rotating}
\usepackage{placeins}

\begin{document}

\title{MCP-DPT: A Defense-Placement Taxonomy and Coverage Analysis for Model Context Protocol Security}

\author{Mehrdad Rostamzadeh}
\email{mrost004@odu.edu}
\orcid{0000-0003-0977-6510}
\affiliation{%
  \institution{Old Dominion University}
  \city{Norfolk}
  \state{Virginia}
  \country{USA}
}

\author{Sidhant Narula}
\affiliation{%
  \institution{Old Dominion University}
  \city{Norfolk}
  \state{Virginia}
  \country{USA}}
\email{snaru004@odu.edu}

\author{Nahom Birhan}
\email{nbirh002@odu.edu}
\affiliation{%
  \institution{Old Dominion University}
  \city{Norfolk}
  \state{Virginia}
  \country{USA}
}

\author{Mohammad Ghasemigol}
\email{mghasemi@odu.edu}
\affiliation{%
 \institution{Old Dominion University}
 \city{Norfolk}
 \state{Virginia}
 \country{USA}}

\author{Daniel Takabi}
\email{takabi@odu.edu}
\affiliation{%
  \institution{Old Dominion Univer}
  \city{Norfolk}
  \state{Virginia}
  \country{USA}}




\renewcommand{\shortauthors}{Rostamzadeh et al.}

\begin{abstract}
The Model Context Protocol (MCP) enables large language models (LLMs)
to dynamically discover and invoke third-party tools, significantly
expanding agent capabilities while introducing a distinct security
landscape. Unlike prompt-only interactions, MCP exposes pre-execution
artifacts, shared context, multi-turn workflows, and third-party supply
chains to adversarial influence across independently operated components.
While recent work has identified MCP-specific attacks and evaluated
defenses, existing studies are largely attack-centric or
benchmark-driven, providing limited guidance on where mitigation
responsibility should reside within the MCP architecture. This is
problematic given MCP's multi-party design and distributed trust
boundaries.
We present a defense-placement--oriented security analysis of MCP,
introducing a layer-aligned taxonomy that organizes attacks by the
architectural component responsible for enforcement. Threats are mapped
across six MCP layers, and primary and secondary defense points are
identified to support principled defense-in-depth reasoning under
adversaries controlling tools, servers, or ecosystem components.
A structured mapping of existing academic and industry defenses onto
this framework reveals uneven and predominantly tool-centric protection,
with persistent gaps at the host orchestration, transport, and
supply-chain layers. These findings suggest that many MCP security
weaknesses stem from architectural misalignment rather than isolated
implementation flaws.
\end{abstract}

\maketitle

\section{Introduction}
Large language models (LLMs) are increasingly deployed as agentic systems
that reason over context~\cite{giurgiu2025supporting}, plan multi-step
actions, and invoke external tools~\cite{wang2025agent} to accomplish user
goals. The Model Context Protocol (MCP) has emerged as a common substrate
for such tool-augmented agents by standardizing how LLM applications connect
to external tools and data sources~\cite{anthropic_mcp_intro,mcp_spec_2025_11_25}.
MCP introduces a host--client--server architecture in which tool providers
expose capabilities via MCP servers, and agent hosts dynamically discover and
invoke those capabilities through standardized messages, schemas, and tool
descriptors~\cite{mcp_spec_2025_11_25}. This design addresses integration
fragmentation and encourages a rapidly growing ecosystem of third-party servers
and reusable connectors~\cite{anthropic_mcp_intro,hou2025mcp_landscape}. At
the same time, MCP shifts the attack surface from a single prompt boundary to
a distributed, multi-party context and tool supply
chain~\cite{errico2025securing}.

A defining characteristic of MCP systems is that they expose rich
pre-execution artifacts---tool names, natural-language descriptions, argument
schemas, and server-provided prompts/resources---that are incorporated into the
model context and directly influence tool selection and
parameterization~\cite{mcp_spec_2025_11_25,zhang2025msb}. Unlike conventional
API integrations, where interfaces are typically static and machine-validated,
MCP explicitly places natural-language metadata in the control loop, creating
new avenues for adversarial influence \emph{before} any tool execution occurs.
Recent measurement work supports this concern at ecosystem scale: large-scale
studies of open-source MCP servers report MCP-specific vulnerabilities beyond
conventional software bugs, including tool poisoning patterns that manifest in
server-provided tool metadata~\cite{hasan2025mcp_first_glance}. More broadly,
MCP surveys highlight security and privacy risks throughout the server lifecycle
(creation, operation, update) and emphasize that distribution and maintenance
practices are inseparable from runtime security for an open and rapidly evolving
protocol ecosystem~\cite{hou2025mcp_landscape}.

Recent work has started to systematize MCP security through dedicated benchmarks
and taxonomies. MCPSecBench~\cite{yang2025mcpsecbench}, MSB~\cite{zhang2025msb},
and MCP-SafetyBench~\cite{zong2025mcpsafetybench} evaluate MCP-specific attacks
across tool-augmented workflows, covering threats that span servers,
hosts/clients, and realistic multi-turn settings. Complementary studies
highlight MCP-native vectors such as tool poisoning and adversarial servers,
demonstrating that malicious behavior can be introduced through tool metadata or
server implementations and propagated through an emerging ecosystem with limited
vetting~\cite{wang2025mcptox,zhao2025when_mcp_servers_attack}.

In parallel, early defense efforts motivate ecosystem-aware mitigations.
SafeMCP~\cite{fang2025safemcp} emphasizes third-party safety risk as a
first-class concern, while MCIP~\cite{jing2025mcip} and
MCP-Guard~\cite{xing2025mcp_guard} propose protocol- and system-level defenses
and accompanying evaluation resources for adversarial testing. These works
collectively suggest that securing MCP-based agents requires controls beyond
prompt filtering, spanning tool-metadata integrity and broader host/client and
ecosystem boundaries.

Despite rapid progress, existing MCP security research remains predominantly
attack-centric: it emphasizes how attacks are executed and how often they
succeed, while offering limited guidance on \emph{where} defenses must be
deployed across MCP's layered trust boundaries and \emph{which stakeholders}
are responsible for enforcing them. This gap is consequential because MCP
deployments are operationally distributed---spanning model provider/alignment,
host/application, client/SDK, server/tool execution, transport/network, and
registry/supply-chain layers---and defenses that are effective in one layer may
be ineffective or unenforceable in
another~\cite{yang2025mcpsecbench,hou2025mcp_landscape}. As a result, current
mitigations tend to cluster around tool-centric or prompt-centric protections,
while structurally critical layers---notably host orchestration, transport, and
supply-chain governance---remain comparatively underdefended.
This paper makes the following contributions to MCP security research.
\begin{enumerate}
  \item We introduce a \emph{defense-placement perspective} for MCP security
        that shifts the focus from attack execution to enforcement responsibility
        across the protocol ecosystem.
  \item We develop a \emph{layer-aligned taxonomy} of MCP-specific attacks that
        reflects real trust boundaries and ownership roles within MCP deployments.
  \item For each attack, we identify \emph{primary} (earliest enforceable) and
        \emph{secondary} (fallback) defense layers, enabling principled
        defense-in-depth reasoning and clearer assignment of mitigation
        responsibility.
  \item We systematically map existing academic and industry defenses onto this
        taxonomy and conduct a \emph{capability-based coverage analysis},
        revealing that current protections are uneven and disproportionately
        tool-centric.
  \item We show that critical layers---particularly transport, host, and
        registry/supply-chain---remain under-defended, indicating that prevailing
        MCP security gaps are \emph{structural} rather than incidental.
\end{enumerate}

\section{Background}

MCP introduces new security risks that extend beyond prompt-level failures and
manifest across clients, protocols, servers, and ecosystem infrastructure.
Existing work has identified many of these attack vectors, but they appear in
different architectural layers and are studied in isolation. This motivates a
structured examination of MCP attack surfaces, prior taxonomies, and their
limitations, as discussed in the following subsections.

\subsection{Tool-Augmented LLMs and MCP}

Modern LLM-based systems increasingly rely on structured interfaces that allow
models to interact with external resources beyond their native
context~\cite{li2023apibank}, extending their functionality through tool
invocation and environment interaction~\cite{yao_react_2023,wang2024llmagentsurvey}.

MCP defines a protocol-level abstraction that enables LLM clients to
dynamically discover, invoke, and interact with tools hosted on heterogeneous
servers~\cite{guo2025systematic,zong2025mcpsafetybench}. By decoupling tool
implementations from model logic, MCP facilitates scalable and interoperable
tool integration, allowing LLM-based agents to operate in distributed
environments with shared context and standardized communication
semantics~\cite{guo2025mcpagentbench,wu2025mcpmark}.

\subsection{MCP Security Implications}

MCP fundamentally reshapes the security landscape of LLM-based
systems~\cite{thirumalaisamy2025aimcp}. Unlike prompt-only deployments, MCP
introduces additional control surface~\cite{hatami2026securing} through
pre-execution artifacts, shared and persistent context, and interactions with
third-party services operating outside the model provider's trust
boundary~\cite{tan2026mcpsandboxscan,Rayarao2025MCP}. Decisions made by LLM
agents are influenced not only by user prompts~\cite{Jishan2024Analyzing}, but
also by tool metadata, schemas, and server-provided context that are
incorporated into the model's reasoning process. These characteristics increase
system expressiveness and interoperability, but also introduce new security
considerations~\cite{maloyan2026promptinjection} that span multiple
architectural layers and stakeholders~\cite{li2025netmcp}, motivating the need
for systematic security analysis beyond traditional prompt-centric
assumptions~\cite{errico2025securing,lei2025mcpsurvey}.

As a consequence, MCP expands the attack surface along three recurring axes.
First, \emph{pre-execution manipulation} of manifests, descriptions, and
schemas can steer agents before any tool invocation; empirical evaluations
demonstrate high attack success from metadata-only tool poisoning in
live-server settings~\cite{wang2025mcptox,guo2025systematic}. Second,
\emph{cross-context propagation} emerges in multi-turn workflows, where
intermediate artifacts, handles, and partial plans enable chained influence
across tools, sessions, and servers; safety degrades as task horizons lengthen
and server interactions accumulate~\cite{zong2025mcpsafetybench,guo2025systematic}.
Third, \emph{third-party control} introduces heterogeneous identity,
authorization, and failure modes outside the model provider's boundary,
motivating ecosystem-level mitigations and protocol-enforceable
controls~\cite{fang2025safemcp,jing2025mcip}.

\subsection{MCP Attack Surfaces and Threat Vectors}

Recent work shows that MCP-related attacks manifest across multiple components
of the system rather than at a single execution point~\cite{guo2025systematic}.
At the MCP client and host level, researchers have identified permission abuse
and excessive capability attacks~\cite{Buhler2025SecuringAgents}, where
over-privileged tool exposure allows agents to perform unintended actions beyond
user intent, particularly in long-running
workflows~\cite{jing2025mcip,errico2025securing}. At the protocol and
context-management layer, context injection and contamination attacks exploit
shared or persistent context to influence downstream reasoning and tool
selection between tasks and sessions~\cite{song2025beyond,zong2025mcpsafetybench}.
Attacks targeting the server and tool execution layer include malicious tool
behavior and covert data exfiltration~\cite{thirumalaisamy2025aimcp}, where
tools return adversarial outputs or leak sensitive information under the guise
of legitimate responses~\cite{fang2025safemcp,guo2025systematic}. In addition,
studies have highlighted registry and supply-chain level risks---such as tool
substitution or version manipulation---which undermine trust assumptions in open
MCP ecosystems~\cite{hasan2025mcp_first_glance,wu2025mcpmark}. While these
attacks demonstrate that vulnerabilities span clients, protocol logic, servers,
and ecosystem infrastructure, existing defenses are typically proposed in
isolation and remain unevenly applied across components, leaving substantial
portions of the MCP attack surface insufficiently covered.

Although several defensive mechanisms have been proposed or deployed in both
academic prototypes~\cite{xing2025mcp_guard,wang2025mindguard,cisco2025mcpscanner,
lasso2025mcpgateway,aim_guard_mcp_2025,eqtylab2025mcpguardian,jing2025mcip}
and industrial MCP implementations, to the best of our knowledge, there is no
existing work that systematically reviews, categorizes, or analyzes MCP defenses
across architectural layers or stakeholders.

\subsection{Prior Work on MCP Security Attack Taxonomies}

Prior research on MCP security has largely focused on enumerating attack vectors
and organizing them into attack-centric taxonomies or benchmark-driven threat
lists.

Song et al.~\cite{song2025beyond} provide one of the earliest structured
analyses of the MCP ecosystem by categorizing malicious server behaviors---such
as tool poisoning, rug-pull attacks, and adversarial external resources---and by
empirically demonstrating their real-world feasibility. Building on this
perspective, Zhao et al.~\cite{zhao2025when_mcp_servers_attack} further
systematize server-side MCP threats by classifying malicious behaviors and
studying their detectability and mitigation challenges.

Several benchmark efforts introduce taxonomies through structured attack lists
and evaluation pipelines. Yang et al.~\cite{yang2025mcpsecbench} propose
MCPSecBench, which organizes MCP attacks across multiple interaction surfaces
and evaluates their success against real-world tools, while Zhang et
al.~\cite{zhang2025msb} introduce MSB, defining attacks along the tool-use
pipeline from planning to execution and response handling. Focusing specifically
on metadata-based manipulation, Wang et al.~\cite{wang2025mcptox} present
MCPTox, demonstrating that high attack success rates can be achieved through
tool poisoning alone without executing malicious tools.

Extending attack categorization to more realistic settings, Zong et
al.~\cite{zong2025mcpsafetybench} introduce MCP-SafetyBench, which classifies
attacks across server-, host-, and user-level interactions and highlights
compounding safety failures in multi-turn, cross-server workflows.
Complementing these studies, the large-scale ecosystem analysis by Hasan et
al.~\cite{hasan2025mcp_first_glance} reveals that MCP-specific vulnerabilities
are widespread in open-source servers, underscoring the systemic nature of the
threat landscape. Finally, Jing et al.~\cite{jing2025mcip} propose MCIP, a
protocol-level safety framework accompanied by a fine-grained taxonomy of unsafe
MCP behaviors, emphasizing contextual integrity rather than isolated exploits.
Collectively, these works establish a rich catalog of MCP attack classes, but
they predominantly organize threats by attack technique or evaluation surface,
offering limited guidance on defense placement across MCP's architectural trust
boundaries.

\subsection{Limitations of Existing MCP Security Research}

Despite recent progress in analyzing the security of the Model Context Protocol
(MCP), existing work exhibits several important limitations.

\begin{itemize}

  \item \textbf{Attack-centric focus.}
        Most studies prioritize identifying attack vectors or measuring
        vulnerability and attack success rates, but do not specify where defenses
        should be deployed within the MCP architecture or which components are
        responsible for enforcement.

  \item \textbf{Lack of defense placement guidance.}
        Existing taxonomies and mitigation efforts rarely distinguish between
        \emph{primary} (earliest feasible) and \emph{secondary} (fallback or
        compensatory) defense layers, making it difficult to reason about defense
        ordering and composition.

  \item \textbf{Partial defense coverage.}
        Proposed defenses---including protocol-level constraints and runtime
        monitoring mechanisms---address only subsets of known attacks, leaving
        gaps at critical layers such as registries, clients, transport, and the
        software supply chain.

  \item \textbf{Unclear defense effectiveness.}
        Current benchmarks primarily evaluate attack feasibility and success, but
        seldom map attacks to available academic or industrial defenses or
        explicitly identify which attacks remain uncovered.

  \item \textbf{Lack of a systematic defense overview.}
        To the best of our knowledge, there is no unified overview or systematic
        review of available MCP defense mechanisms across academic and industrial
        deployments, making it difficult to assess defense coverage, compare
        approaches, or reason about residual risk at the ecosystem level.

\end{itemize}


Table~\ref{tab:mcp_compare} compares six representative MCP security works
against our taxonomy across nine evaluation dimensions. The first two rows
show that all prior works adopt either an attack-surface, pipeline-stage, or
scenario-based taxonomy type --- none organizes threats by where defenses must
be enforced. Rows three through five reveal that while attack-centricity is
universal ($\checkmark$ across all six works), control- and defense-centric
reasoning is only partial ($\sim$) in MCPSecBench and When MCP Servers Attack,
and absent in MSB, MCP-SafetyBench, and MCPLib. Critically, no prior work
explicitly maps attacks to defense layers or assigns enforcement responsibility
to specific architectural stakeholders --- both dimensions receive $\times$
across all six columns. Trust-boundary awareness and defense-in-depth reasoning
appear sporadically, with MCIP being the strongest prior work ($\checkmark$ on
defense-in-depth), yet even MCIP does not identify primary versus secondary
defense points or expose under-defended layers. The final three rows ---
primary/secondary defense points, identifying under-defended layers, and
coverage analysis --- are exclusive to this work, marking the core contribution
of MCP-DPT: shifting the framing from \emph{what attacks exist} to
\emph{where and by whom they must be stopped}.

\FloatBarrier

\section{Taxonomy}
\begin{figure}
  \centering
  \includegraphics[width=0.85\textwidth]{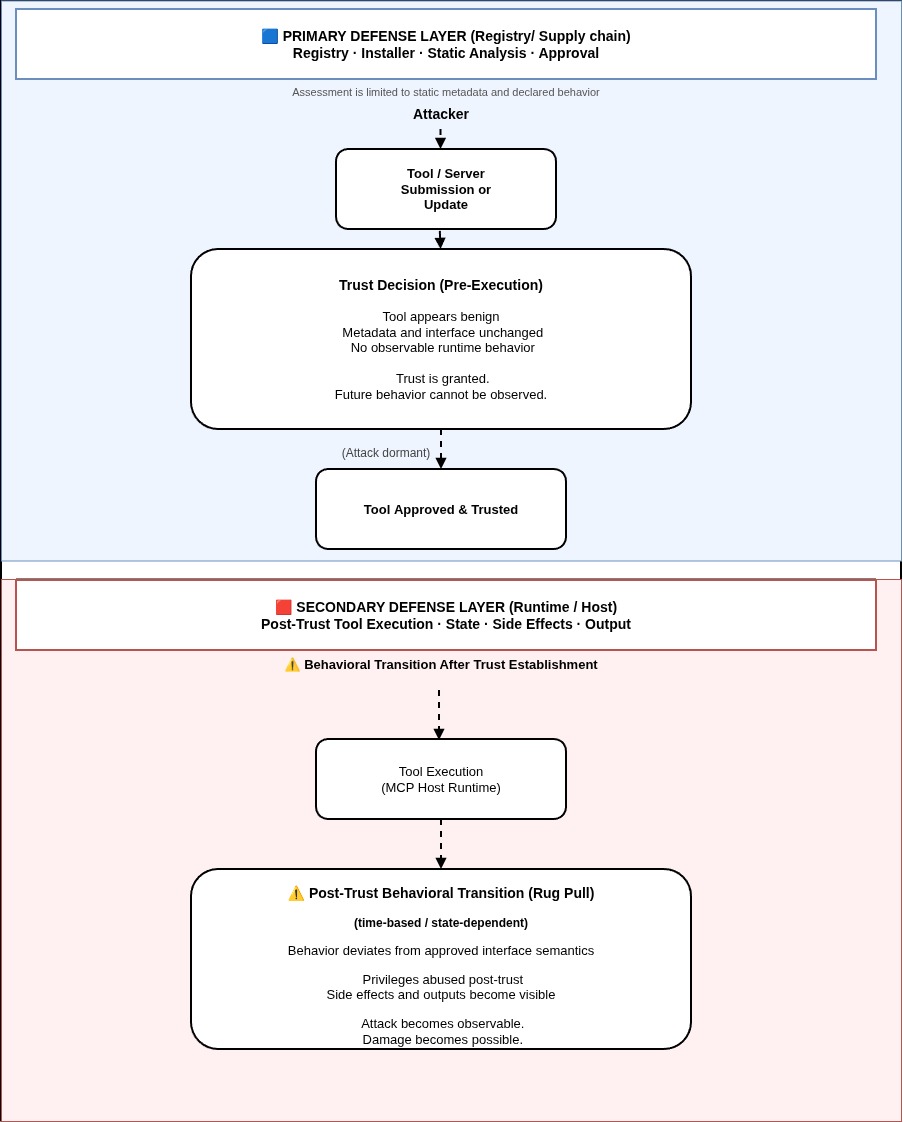}
  \caption{Primary and secondary defense layers for a rug pull attack.}
  \label{fig:rugpull_attack}
\end{figure}

We conducted a structured review of MCP security literature published through 2025 and 2026, collecting attack descriptions from benchmarks \cite{yang2025mcpsecbench, guo2025systematic, zhao2025when_mcp_servers_attack}, server-side threat analyses \cite{zhao2025when_mcp_servers_attack, song2025beyond, stacklok2025toolhive}, ecosystem measurement studies \cite{hasan2025mcp_first_glance}, and protocol safety frameworks \cite{jing2025mcip}. Each identified attack was assigned to the MCP architectural layer first becomes enforceable. We then assigned a primary defense layer—the earliest boundary where prevention is feasible with sufficient authority and visibility—and a secondary defense layer—the fallback containment point if the primary fails. This assignment was guided by the trust boundaries and ownership roles inherent in each layer \cite{hou2025mcp_landscape} not by the frequency of attacks or empirical success rates.

\begin{figure*}[!tb]
  \centering
  \includegraphics[width=1\textwidth]{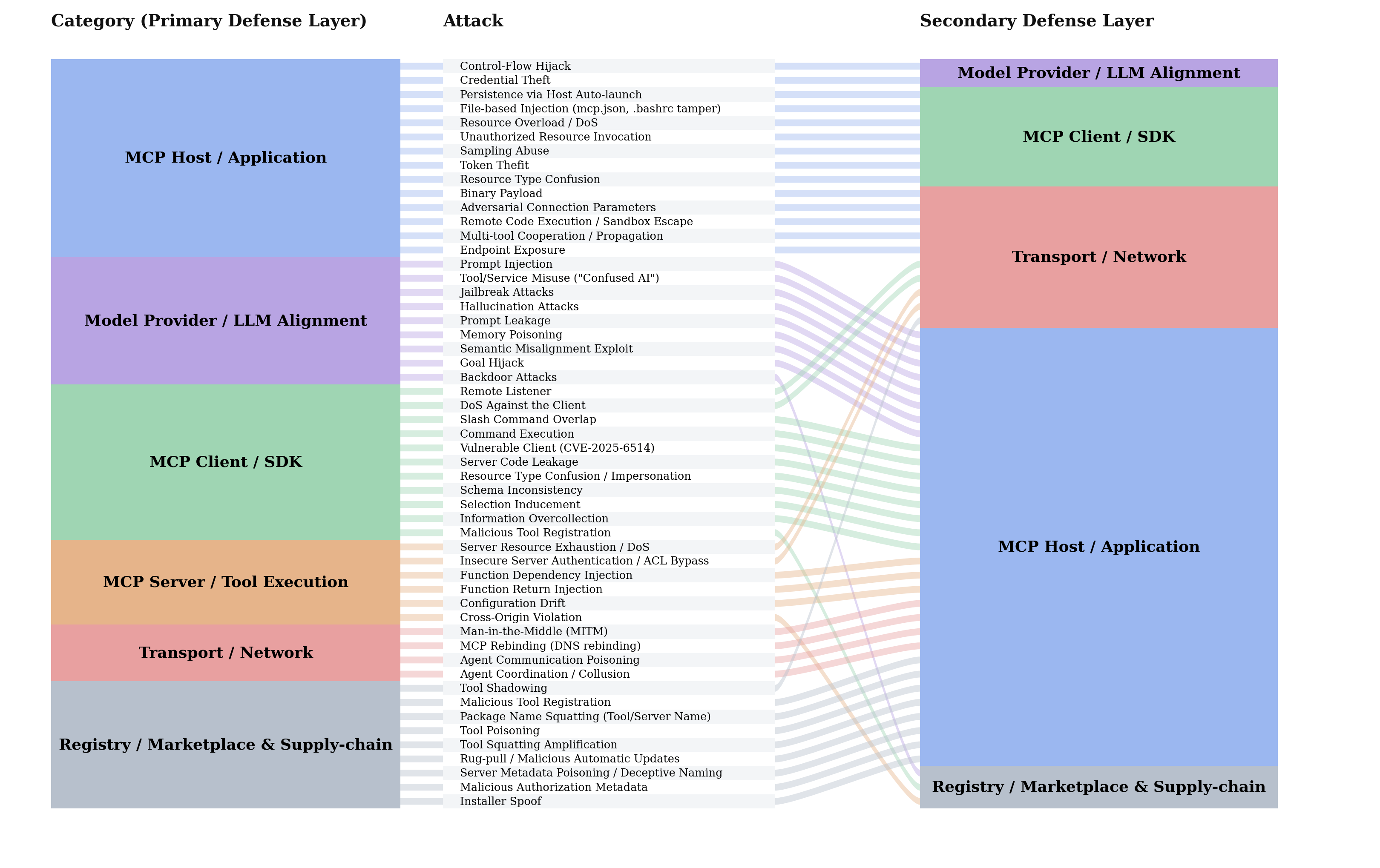}
  \caption{Defense placement taxonomy of attacks.}
  \label{fig:taxonomy}
\end{figure*}

In our defense-placement-oriented taxonomy for MCP
security, we classify attacks based on where countermeasures must be
implemented, rather than by attack technique or attacker intent as in previous
work~\cite{yang2025mcpsecbench,zhao2025when_mcp_servers_attack}. Our framework
organizes threats according to the architectural layer primarily responsible for
mitigation---Model Provider/LLM Alignment, MCP Host/Application, MCP
Client/SDK, MCP Server/Tool Execution, Transport/Network, and
Registry/Marketplace \& Supply-chain---reflecting the ownership and trust
boundaries of the MCP ecosystem. By explicitly aligning each attack with the
component that must enforce its defense, this perspective yields a more
actionable understanding of vulnerabilities, clarifies the boundaries of
responsibility, and enables more effective and targeted mitigation strategies.
As shown in Figure~\ref{fig:taxonomy}, every attack entry is associated with a
primary and secondary defense layer, guiding practitioners toward an effective
placement of safeguards. To our knowledge, this is the first taxonomy that
structures MCP threats around defense placement, providing a practical,
deployment-oriented framework for securing MCP-based systems.

%

\begin{table}[htbp]
\centering
\caption{Comparison of MCP security benchmarks and analyses with respect to
         defense placement and enforcement responsibility.
         $\checkmark$~=~fully supported;\enspace
         $\sim$~=~partially supported;\enspace
         $\times$~=~not supported.}
\label{tab:mcp_compare}
\resizebox{\linewidth}{!}{%
\renewcommand{\arraystretch}{1.3}
\begin{tabular}{@{}lccccccc@{}}
\toprule
\textbf{Aspect}
  & \textbf{\shortstack{MCPSec\\Bench~\cite{yang2025mcpsecbench}}}
  & \textbf{\shortstack{MSB\\\cite{zhang2025msb}}}
  & \textbf{\shortstack{MCP-Safety\\Bench~\cite{zong2025mcpsafetybench}}}
  & \textbf{\shortstack{MCPLib\\\cite{guo2025systematic}}}
  & \textbf{\shortstack{When MCP Servers\\Attack~\cite{zhao2025when_mcp_servers_attack}}}
  & \textbf{\shortstack{MCIP\\\cite{jing2025mcip}}}
  & \textbf{MCP-DPT (This Work)} \\
\midrule
Primary focus
  & Security benchmark
  & Attack benchmark
  & Safety benchmark
  & \shortstack{Attack library/\\taxonomy}
  & \shortstack{Malicious MCP\\servers}
  & \shortstack{Safety-enhanced\\protocol}
  & Defense placement \\[4pt]
Taxonomy type
  & Attack-surface
  & \shortstack{Pipeline-stage\\attack}
  & Safety-scenario
  & \shortstack{Attack-method\\taxonomy}
  & \shortstack{Server-component\\attack}
  & \shortstack{Unsafe-behavior/\\lifecycle taxonomy}
  & \shortstack{Defense-placement\\taxonomy} \\[4pt]
Attack-centric
  & $\checkmark$ & $\checkmark$ & $\checkmark$ & $\checkmark$ & $\checkmark$ & $\sim$ & $\times$ \\
Control-/defense-centric
  & $\sim$ & $\times$ & $\times$ & $\times$ & $\sim$ & $\checkmark$ & $\checkmark$ \\
Maps attacks $\rightarrow$ defenses
  & $\times$ & $\times$ & $\times$ & $\times$ & $\sim$ & $\sim$ & $\checkmark$ \\
Assigns enforcement responsibility
  & $\times$ & $\times$ & $\times$ & $\times$ & $\times$ & $\times$ & $\checkmark$ \\
Trust-boundary aware
  & $\sim$ & $\times$ & $\sim$ & $\times$ & $\sim$ & $\sim$ & $\checkmark$ \\
Defense-in-depth reasoning
  & $\sim$ & $\times$ & $\times$ & $\times$ & $\sim$ & $\checkmark$ & $\checkmark$ \\
Primary/secondary defense points
  & $\times$ & $\times$ & $\times$ & $\times$ & $\times$ & $\times$ & $\checkmark$ \\
Identifies under-defended layers
  & $\times$ & $\times$ & $\times$ & $\times$ & $\times$ & $\times$ & $\checkmark$ \\
Supports coverage analysis
  & $\times$ & $\times$ & $\times$ & $\times$ & $\times$ & $\times$ & $\checkmark$ \\
\bottomrule
\end{tabular}%
}
\end{table}

The taxonomy is structured around three conceptual axes: (i)~the
\emph{primary defense layer} (where the attack must first be prevented),
(ii)~the \emph{attack class} itself, and (iii)~the \emph{secondary defense
layer} (where residual risk must be mitigated if the primary defense fails).
This structure shifts the focus from how attacks are executed to where defenses
must be deployed within the MCP ecosystem.
The appendix provides detailed descriptions and illustrative examples for a
representative subset of 49~attacks that cover all architectural layers and
attack classes.

\subsection{Architectural Layers as Security Boundaries}

The taxonomy is grounded in the MCP architecture and its real operational trust
boundaries. We define six layers, each corresponding to a distinct ownership
domain and control surface.

\subsubsection{Model Provider / LLM Alignment}

This layer encompasses the language model itself and the mechanisms that govern
its behavior, including alignment policies, refusal logic, tool selection
behavior, and training data assumptions, which are typically established through
alignment techniques such as reinforcement learning from human feedback
(RLHF)~\cite{ouyang2022training}. It is responsible for ensuring that model
output remains semantically safe, policy-compliant, and aligned with the intent
of the developer. Security failures at this layer typically stem from
limitations in model reasoning, degradation of alignment during fine-tuning
~\cite{yang2025alleviating}, or compromised training integrity---including the
presence of latent backdoors that can selectively bypass alignment constraints
at inference time~\cite{shen2025bait}.

\subsubsection{MCP Host / Application}

The host layer occupies a central position in the MCP architecture because it
is responsible for mediating between model outputs and real-world tool
execution. As the component that embeds the language model within an application
context, the host governs execution state, capability exposure, resource
boundaries, and orchestration logic across tools and services. Prior work has
shown that many failures in LLM-based systems arise not from the model itself,
but from insufficient enforcement at the application boundary where model
decisions are translated into actions~\cite{wei2023jailbroken,kang2024exploiting}.
Because the host layer retains visibility into tool schemas, execution intent,
and system state, it provides a uniquely effective location to enforce safety
invariants, validate model outputs, and constrain behavior before irreversible
side effects occur~\cite{guo2025systematic}. For these reasons, the host layer
forms a distinct and indispensable element of the taxonomy, reflecting both its
architectural responsibility and its role as the final programmable control
point prior to tool invocation.

\subsubsection{MCP Client / SDK}

The MCP Client/SDK layer comprises the software runtime that bridges the
language model-driven agent with external MCP servers, handling protocol
parsing, request construction, response interpretation, and local execution
logic. This layer operates at a sensitive trust boundary, as it transforms
high-level model decisions into concrete network interactions and local actions,
often within proximity to user resources and
credentials~\cite{hou2025mcp_landscape}. Security risks arise when client
implementations implicitly trust server responses, fail to enforce strict
schemas, or allow model-influenced output to trigger unintended behavior.
Because the client mediates both inbound data from servers and outbound actions
initiated by the model, weaknesses at this layer can amplify upstream failures
or enable direct compromise even when other components behave correctly.
Consequently, the client layer is a critical enforcement point for validation,
isolation, and permission scope within the MCP
architecture~\cite{xing2025mcp_guard}.

\subsubsection{MCP Server / Tool Execution}

The MCP Server/Tool Execution layer represents the runtime environment in which
MCP tools are implemented and executed~\cite{tan2026mcpsandboxscan,Rayarao2025MCP},
including server-side logic, exposed APIs, authentication checks, plugin
loading, and resource management. This layer is responsible for safely handling
inputs received from MCP hosts and clients, enforcing isolation boundaries, and
ensuring that tool execution does not violate system or data
integrity~\cite{yang2025mcpsecbench}. Security failures here typically arise
from unsafe execution environments, weak authentication or authorization, overly
permissive introspection interfaces, or inadequate isolation between tools and
underlying infrastructure. Because this layer directly executes code and
processes data on behalf of the LLM, it forms a critical containment boundary:
once compromised, attacks can escalate beyond the MCP ecosystem into broader
systems, making robust server-side safeguards essential for defense in
depth~\cite{thirumalaisamy2025aimcp}.

\subsubsection{Transport / Network}

The Transport/Network layer captures the communication channels over which MCP
messages, tool invocations, and responses are exchanged between hosts, clients,
and servers. This layer encompasses network protocols, session management,
message framing, encryption, authentication of endpoints, and guarantees related
to integrity, freshness, and ordering. Unlike higher layers that reason about
intent or semantics, the transport layer is responsible for ensuring that
messages are delivered faithfully and only between authenticated parties, even
in the presence of network-level adversaries.

Security failures at this layer arise from weak or absent transport protections,
such as insufficient authentication of MCP endpoints, lack of message integrity
or replay protection, improper session binding, or insecure transport
configurations. These weaknesses can enable man-in-the-middle attacks, replay
or reordering of tool calls, downgrade attacks, or injection of stale or
malicious responses that remain syntactically valid at higher layers. Because
MCP agents often assume reliable and trustworthy communication, transport-layer
compromises can silently subvert otherwise correct behavior at the client, host,
or server layers, making robust network-level safeguards essential for
preserving end-to-end security
guarantees~\cite{yang2025mcpsecbench,hasan2025mcp_first_glance}.

\subsubsection{Registry / Marketplace \& Supply-chain}

The Registry/Marketplace--Supply-chain layer captures all mechanisms responsible
for discovery, distribution, versioning, and provenance of MCP servers, tools,
plugins, and their updates. This layer sits upstream of execution and governs
how components enter the MCP ecosystem in the first place. Its security role is
to ensure authenticity, integrity, and traceability of MCP artifacts before they
are ever invoked by a host or client. Failures at this layer undermine trust
assumptions across the entire stack: once a malicious or compromised component
is registered, updated, or promoted through the supply chain, downstream layers
may execute it correctly yet still produce unsafe outcomes. As a result, this
layer is fundamental for enforcing trust boundaries, provenance guarantees, and
long-term ecosystem safety rather than runtime correctness
alone~\cite{hasan2025mcp_first_glance}.


\subsection{Attack}
This dimension captures the specific malicious technique or exploit employed by
an adversary, describing the mechanism by which the attack is realized within
the MCP ecosystem. Rather than exhaustively defining individual attacks in the
main text, we use this dimension to reference established MCP-specific threat
types and illustrate how they manifest across architectural layers. Detailed
definitions and assumptions for each attack type are
provided in Appendix \ref{app:attacks}, allowing the core taxonomy to focus on structural
relationships between attacks, defenses, and enforcement responsibilities.

\subsection{Primary Defense Layer}

The primary defense layer refers to the first line of protection designed to
prevent an attack from occurring in the first place, or to block it at the
earliest possible point of interaction with the system. This layer operates
closest to the attack source, aiming to stop malicious inputs, behaviors, or
configurations before they can influence downstream components.
In the context of MCP, this notion is especially important because attacks may originate through different entry points—such as model-facing metadata, client interactions, server behavior, transport channels, or supply-chain artifacts—while their effects may only become visible later in the execution flow. For this reason, the primary defense layer is not simply the place where an attack is eventually detected, but the earliest architectural boundary where meaningful prevention can be enforced with sufficient authority and visibility.

By identifying this earliest enforceable boundary, the primary defense layer provides a principled way to assign mitigation responsibility across the MCP ecosystem. It clarifies which component should serve as the first point of intervention before malicious influence propagates into later stages such as orchestration, tool invocation, or runtime execution. This is particularly valuable in a distributed architecture like MCP, where different layers possess different control surfaces and trust assumptions. Accordingly, the primary defense layer serves as the foundation for defense-in-depth: it is responsible for blocking the threat at entry, while downstream layers act as secondary safeguards to contain impact if that initial protection fails.

\subsection{Secondary Defense Layer}

The secondary defense layer provides defense-in-depth by limiting the impact
and propagation of an attack when the primary defense layer fails, is
misconfigured, or is deliberately bypassed. Rather than preventing initial
compromise, this layer focuses on containment, mitigation, and recovery.

To clarify how primary and secondary defense layers differ in practice,
Figure~\ref{fig:rugpull_attack} presents a concrete example based on a rug pull
attack. The diagram illustrates how a malicious MCP tool can successfully pass
pre-execution trust decisions at the registry and approval stage---where
evaluation is limited to static metadata, declared interfaces, and expected
behavior---while deferring its malicious actions until after trust has been
established. Once the trusted tool is executed at runtime, its behavior
transitions in a time- or state-dependent manner, making the attack observable
only within the secondary (host-level) defense layer. This example is not
intended to propose a mitigation strategy, but rather to illustrate how defense
responsibilities are distributed across layers in the taxonomy.

\subsection{Motivation for Defense Layer Stratification
}
The primary/secondary distinction was motivated by two observations:

First, MCP is a distributed, multi-owner architecture \cite{hou2025mcp_landscape} in which no single component controls the entire stack\cite{errico2025securing}, making a single defense point inherently insufficient. An attack may enter at one layer but only become observable or stoppable at another — as Figure~\ref{fig:rugpull_attack} illustrates. The registry layer, for instance, can only evaluate static metadata at submission time, whereas malicious behavior may only manifest at runtime within the host layer\cite{song2025beyond}.

Second, existing defenses fail silently at layer boundaries. As shown in Table~\ref{tab:mcp_compare}, no existing taxonomy assigns enforcement responsibility to specific architectural components. Consequently, practitioners deploying a single tool-layer defense have no guidance on what happens if that defense fails, is misconfigured, or is deliberately bypassed\cite{xing2025mcp_guard, jing2025mcip}.

\section{Available Defense Mechanisms}
This section provides an overview of existing defense mechanisms specifically
designed to address attacks in MCP-based systems. The goal is to review their
key features and to analyze which classes of MCP attacks they are capable of
mitigating.

In this work, the effectiveness of a defense mechanism is assessed in terms of
\emph{capability-based coverage} rather than empirical detection accuracy. A
defense is considered to cover an attack class if it can reasonably detect,
block, or constrain that class under commonly assumed MCP threat models and
deployment conditions. This notion of coverage reflects whether a defense is
architecturally positioned to mitigate an attack type, not whether it achieves
perfect prevention in practice. Consequently, our analysis focuses on structural
defense applicability and enforcement potential rather than quantitative metrics.
Tables~\ref{tab:heatmap_full}
illustrates how each
attack is addressed by the available defense mechanisms described in the
following sections.

\begin{table*}[t] 
\centering
\caption{Capability-based defense coverage across all 49 taxonomy attacks and 13 defense mechanisms, organized by the primary MCP architectural layer responsible for enforcement. $\bullet$~=~covered;\enspace $\circ$~=~not covered.}
\label{tab:heatmap_full}
\vspace{0.5em}

\setlength{\tabcolsep}{3pt}
\renewcommand{\arraystretch}{0.85}

\resizebox{\linewidth}{!}{%
\begin{tabular}{@{}lccccccccccccc@{}}
\toprule
\textbf{Attack}
  & \rotatebox{66}{\textbf{MCP-Scan}}
  & \rotatebox{66}{\textbf{MCIP-Guardian}}
  & \rotatebox{66}{\textbf{Cisco MCP Scanner}}
  & \rotatebox{66}{\textbf{MCP-Gateway}}
  & \rotatebox{66}{\textbf{MCP Guardian (eqtylab)}}
  & \rotatebox{66}{\textbf{MCP-Defender}}
  & \rotatebox{66}{\textbf{MCP-Guard}}
  & \rotatebox{66}{\textbf{Prisma AIRS}}
  & \rotatebox{66}{\textbf{MCPScan.ai}}
  & \rotatebox{66}{\textbf{MCP-Shield}}
  & \rotatebox{66}{\textbf{ToolHive}}
  & \rotatebox{66}{\textbf{MindGuard}}
  & \rotatebox{66}{\textbf{AIM-Guard-MCP}} \\
\midrule
\multicolumn{14}{@{}l}{\textbf{Model Provider / LLM Alignment}} \\
\midrule
Prompt Injection                        & $\bullet$ & $\bullet$ & $\bullet$ & $\circ$ & $\circ$ & $\bullet$ & $\bullet$ & $\bullet$ & $\circ$ & $\bullet$ & $\circ$ & $\circ$ & $\bullet$ \\
Tool/Service Misuse (``Confused AI'')   & $\circ$ & $\bullet$ & $\circ$ & $\circ$ & $\bullet$ & $\bullet$ & $\bullet$ & $\bullet$ & $\circ$ & $\circ$ & $\circ$ & $\circ$ & $\bullet$ \\
Jailbreak Attacks                       & $\circ$ & $\circ$ & $\circ$ & $\circ$ & $\circ$ & $\circ$ & $\bullet$ & $\bullet$ & $\circ$ & $\circ$ & $\circ$ & $\circ$ & $\circ$ \\
Hallucination Attacks                   & $\circ$ & $\bullet$ & $\circ$ & $\circ$ & $\circ$ & $\circ$ & $\bullet$ & $\bullet$ & $\circ$ & $\circ$ & $\circ$ & $\circ$ & $\circ$ \\
Prompt Leakage                          & $\bullet$ & $\bullet$ & $\bullet$ & $\circ$ & $\circ$ & $\circ$ & $\bullet$ & $\bullet$ & $\circ$ & $\bullet$ & $\circ$ & $\circ$ & $\circ$ \\
Memory Poisoning                        & $\circ$ & $\circ$ & $\circ$ & $\circ$ & $\circ$ & $\circ$ & $\circ$ & $\circ$ & $\circ$ & $\circ$ & $\circ$ & $\bullet$ & $\circ$ \\
Semantic Misalignment Exploit           & $\circ$ & $\circ$ & $\circ$ & $\circ$ & $\circ$ & $\circ$ & $\circ$ & $\circ$ & $\circ$ & $\circ$ & $\circ$ & $\circ$ & $\circ$ \\
Goal Hijack                             & $\circ$ & $\circ$ & $\circ$ & $\circ$ & $\circ$ & $\circ$ & $\circ$ & $\circ$ & $\circ$ & $\circ$ & $\circ$ & $\circ$ & $\circ$ \\
Backdoor Attacks                        & $\circ$ & $\bullet$ & $\circ$ & $\circ$ & $\bullet$ & $\circ$ & $\bullet$ & $\circ$ & $\circ$ & $\circ$ & $\circ$ & $\circ$ & $\circ$ \\
\midrule
\multicolumn{14}{@{}l}{\textbf{MCP Server / Tool Execution}} \\
\midrule
Server Resource Exhaustion / DoS        & $\circ$ & $\circ$ & $\circ$ & $\circ$ & $\circ$ & $\circ$ & $\bullet$ & $\circ$ & $\circ$ & $\circ$ & $\circ$ & $\circ$ & $\circ$ \\
Insecure Server Authentication          & $\circ$ & $\circ$ & $\circ$ & $\circ$ & $\bullet$ & $\circ$ & $\bullet$ & $\circ$ & $\circ$ & $\circ$ & $\circ$ & $\circ$ & $\circ$ \\
Function Dependency Injection           & $\bullet$ & $\circ$ & $\circ$ & $\circ$ & $\bullet$ & $\circ$ & $\bullet$ & $\bullet$ & $\circ$ & $\circ$ & $\circ$ & $\circ$ & $\circ$ \\
Function Return Injection               & $\bullet$ & $\bullet$ & $\circ$ & $\circ$ & $\bullet$ & $\bullet$ & $\bullet$ & $\bullet$ & $\circ$ & $\circ$ & $\circ$ & $\circ$ & $\circ$ \\
Configuration Drift                     & $\circ$ & $\bullet$ & $\circ$ & $\circ$ & $\bullet$ & $\circ$ & $\bullet$ & $\circ$ & $\circ$ & $\circ$ & $\circ$ & $\circ$ & $\circ$ \\
Cross-Origin Violation                  & $\circ$ & $\bullet$ & $\circ$ & $\bullet$ & $\bullet$ & $\circ$ & $\bullet$ & $\circ$ & $\circ$ & $\circ$ & $\circ$ & $\circ$ & $\circ$ \\
\midrule
\multicolumn{14}{@{}l}{\textbf{Transport / Network}} \\
\midrule
Man-in-the-Middle (MiTM)                & $\circ$ & $\circ$ & $\circ$ & $\bullet$ & $\circ$ & $\circ$ & $\circ$ & $\circ$ & $\circ$ & $\circ$ & $\circ$ & $\circ$ & $\circ$ \\
MCP Rebinding (DNS Rebinding)           & $\circ$ & $\circ$ & $\circ$ & $\bullet$ & $\circ$ & $\circ$ & $\circ$ & $\circ$ & $\circ$ & $\circ$ & $\circ$ & $\circ$ & $\circ$ \\
Agent Communication Poisoning           & $\circ$ & $\circ$ & $\circ$ & $\circ$ & $\circ$ & $\circ$ & $\circ$ & $\circ$ & $\circ$ & $\circ$ & $\circ$ & $\circ$ & $\circ$ \\
Agent Coordination / Collusion          & $\circ$ & $\circ$ & $\circ$ & $\bullet$ & $\circ$ & $\circ$ & $\circ$ & $\circ$ & $\circ$ & $\circ$ & $\circ$ & $\circ$ & $\circ$ \\
\midrule
\multicolumn{14}{@{}l}{\textbf{Registry / Marketplace \& Supply-Chain}} \\
\midrule
Tool Shadowing (Mirror Attacks)         & $\bullet$ & $\bullet$ & $\circ$ & $\circ$ & $\bullet$ & $\circ$ & $\bullet$ & $\circ$ & $\bullet$ & $\bullet$ & $\circ$ & $\circ$ & $\circ$ \\
Malicious Tool Registration             & $\bullet$ & $\circ$ & $\circ$ & $\circ$ & $\circ$ & $\circ$ & $\bullet$ & $\circ$ & $\bullet$ & $\bullet$ & $\circ$ & $\circ$ & $\circ$ \\
Package Name Squatting (Tool / Server)  & $\bullet$ & $\bullet$ & $\circ$ & $\circ$ & $\circ$ & $\circ$ & $\bullet$ & $\circ$ & $\bullet$ & $\bullet$ & $\circ$ & $\circ$ & $\circ$ \\
Tool Poisoning Attack                   & $\bullet$ & $\bullet$ & $\bullet$ & $\circ$ & $\bullet$ & $\bullet$ & $\bullet$ & $\bullet$ & $\bullet$ & $\bullet$ & $\circ$ & $\bullet$ & $\bullet$ \\
Tool Squatting Amplification            & $\bullet$ & $\bullet$ & $\circ$ & $\circ$ & $\circ$ & $\circ$ & $\bullet$ & $\circ$ & $\bullet$ & $\circ$ & $\circ$ & $\circ$ & $\circ$ \\
Promotional / Deceptive Metadata        & $\circ$ & $\circ$ & $\circ$ & $\circ$ & $\circ$ & $\circ$ & $\circ$ & $\circ$ & $\circ$ & $\circ$ & $\circ$ & $\circ$ & $\circ$ \\
\midrule
\multicolumn{14}{@{}l}{\textbf{MCP Host / Application}} \\
\midrule
Control-Flow Hijack                     & $\circ$ & $\circ$ & $\circ$ & $\circ$ & $\circ$ & $\circ$ & $\circ$ & $\circ$ & $\circ$ & $\circ$ & $\circ$ & $\circ$ & $\circ$ \\
Credential Theft                        & $\circ$ & $\circ$ & $\circ$ & $\circ$ & $\circ$ & $\circ$ & $\circ$ & $\circ$ & $\circ$ & $\circ$ & $\circ$ & $\circ$ & $\circ$ \\
Persistence via Host Auto-Launch        & $\circ$ & $\bullet$ & $\circ$ & $\circ$ & $\bullet$ & $\circ$ & $\circ$ & $\circ$ & $\circ$ & $\circ$ & $\bullet$ & $\circ$ & $\circ$ \\
File-Based Injection                    & $\bullet$ & $\bullet$ & $\bullet$ & $\circ$ & $\circ$ & $\circ$ & $\bullet$ & $\bullet$ & $\circ$ & $\bullet$ & $\circ$ & $\circ$ & $\circ$ \\
Resource Overload / DoS                 & $\circ$ & $\circ$ & $\circ$ & $\circ$ & $\circ$ & $\circ$ & $\circ$ & $\circ$ & $\circ$ & $\circ$ & $\circ$ & $\circ$ & $\circ$ \\
Unauthorized Resource Invocation        & $\circ$ & $\bullet$ & $\circ$ & $\circ$ & $\bullet$ & $\circ$ & $\circ$ & $\circ$ & $\circ$ & $\circ$ & $\circ$ & $\circ$ & $\circ$ \\
Sampling Abuse                          & $\circ$ & $\circ$ & $\circ$ & $\circ$ & $\bullet$ & $\bullet$ & $\circ$ & $\bullet$ & $\circ$ & $\circ$ & $\circ$ & $\circ$ & $\circ$ \\
Token Theft                             & $\circ$ & $\circ$ & $\circ$ & $\circ$ & $\circ$ & $\circ$ & $\circ$ & $\circ$ & $\circ$ & $\circ$ & $\circ$ & $\circ$ & $\circ$ \\
Resource Type Confusion                 & $\circ$ & $\circ$ & $\circ$ & $\circ$ & $\circ$ & $\circ$ & $\circ$ & $\circ$ & $\circ$ & $\circ$ & $\circ$ & $\circ$ & $\circ$ \\
Binary Payload Delivery                 & $\circ$ & $\circ$ & $\circ$ & $\circ$ & $\circ$ & $\circ$ & $\circ$ & $\circ$ & $\circ$ & $\circ$ & $\circ$ & $\circ$ & $\circ$ \\
Remote Code Execution                   & $\circ$ & $\circ$ & $\circ$ & $\circ$ & $\circ$ & $\circ$ & $\bullet$ & $\circ$ & $\circ$ & $\circ$ & $\circ$ & $\circ$ & $\circ$ \\
Multi-Tool Cooperation / Propagation    & $\circ$ & $\circ$ & $\circ$ & $\circ$ & $\circ$ & $\circ$ & $\circ$ & $\circ$ & $\circ$ & $\circ$ & $\circ$ & $\circ$ & $\circ$ \\
Endpoint Exposure                       & $\circ$ & $\circ$ & $\circ$ & $\circ$ & $\circ$ & $\bullet$ & $\circ$ & $\circ$ & $\circ$ & $\circ$ & $\circ$ & $\circ$ & $\circ$ \\
\midrule
\multicolumn{14}{@{}l}{\textbf{MCP Client / SDK}} \\
\midrule
Remote Listener                         & $\bullet$ & $\bullet$ & $\circ$ & $\circ$ & $\bullet$ & $\bullet$ & $\bullet$ & $\bullet$ & $\circ$ & $\bullet$ & $\circ$ & $\circ$ & $\circ$ \\
DoS Against the Client                  & $\circ$ & $\circ$ & $\circ$ & $\circ$ & $\circ$ & $\circ$ & $\circ$ & $\circ$ & $\circ$ & $\circ$ & $\circ$ & $\circ$ & $\circ$ \\
Slash Command Overlap                   & $\circ$ & $\circ$ & $\circ$ & $\circ$ & $\circ$ & $\circ$ & $\circ$ & $\circ$ & $\circ$ & $\circ$ & $\circ$ & $\circ$ & $\circ$ \\
Command Execution                       & $\circ$ & $\circ$ & $\circ$ & $\circ$ & $\circ$ & $\circ$ & $\bullet$ & $\circ$ & $\circ$ & $\circ$ & $\circ$ & $\circ$ & $\circ$ \\
Vulnerable Client (CVE-2025-6514)       & $\circ$ & $\circ$ & $\circ$ & $\circ$ & $\circ$ & $\circ$ & $\bullet$ & $\circ$ & $\circ$ & $\circ$ & $\circ$ & $\circ$ & $\bullet$ \\
Server Code Leakage                     & $\bullet$ & $\circ$ & $\circ$ & $\circ$ & $\circ$ & $\circ$ & $\circ$ & $\circ$ & $\circ$ & $\circ$ & $\circ$ & $\circ$ & $\circ$ \\
Resource Type Confusion / Impersonation & $\circ$ & $\circ$ & $\circ$ & $\circ$ & $\circ$ & $\circ$ & $\circ$ & $\circ$ & $\circ$ & $\circ$ & $\circ$ & $\circ$ & $\circ$ \\
Schema Inconsistency                    & $\circ$ & $\bullet$ & $\circ$ & $\circ$ & $\circ$ & $\circ$ & $\bullet$ & $\circ$ & $\circ$ & $\circ$ & $\circ$ & $\circ$ & $\circ$ \\
Selection Inducement                    & $\bullet$ & $\bullet$ & $\bullet$ & $\circ$ & $\circ$ & $\circ$ & $\bullet$ & $\bullet$ & $\circ$ & $\bullet$ & $\circ$ & $\circ$ & $\circ$ \\
Information Overcollection              & $\circ$ & $\bullet$ & $\circ$ & $\circ$ & $\bullet$ & $\bullet$ & $\circ$ & $\bullet$ & $\circ$ & $\circ$ & $\circ$ & $\circ$ & $\circ$ \\
Malicious Tool Installation             & $\bullet$ & $\circ$ & $\circ$ & $\circ$ & $\circ$ & $\circ$ & $\circ$ & $\circ$ & $\bullet$ & $\bullet$ & $\circ$ & $\circ$ & $\circ$ \\
\bottomrule
\end{tabular}%
} 
\end{table*}

\subsection{Static / Pre-Execution Defenses}

Static/pre-execution defenses analyze MCP configurations, tool metadata,
manifests, and policies before execution to identify security risks without
observing runtime behavior~\cite{jamshidi2025securingmcp}.

\subsection{Behavior-Level / Runtime Defenses}

Behavior-level/runtime defenses monitor and analyze the runtime behavior of
LLMs and tool interactions, focusing on detecting anomalous, malicious, or
policy-violating actions during execution.
\subsection{Isolation-Based / Architecture Defenses}

Isolation-based/architecture defenses redesign the system architecture to
enforce strict trust boundaries between LLMs, tools, and servers, limiting the
impact of compromised components~\cite{xing2025mcp_guard}.

\subsection{Decision-Level Defenses}

Decision-level defenses protect the LLM's internal decision-making process,
particularly tool selection and parameterization, by analyzing how decisions are
formed rather than only their final outcomes~\cite{wang2025mindguard}. These
approaches monitor intermediate decision signals to identify subtle manipulation
attempts---such as tool poisoning or preference steering---that may evade
output-level defenses. By intervening during decision-making, they can detect
attacks before harmful actions are executed. This capability is especially
important in multi-step, context-rich MCP workflows. As a result, decision-level
defenses complement traditional outcome-based protections.

\subsection{Defense Mechanisms}

\subsubsection{MCP-Scan (Invariant Labs)}

MCP-Scan is a static security analysis tool that inspects MCP client
configurations and tool metadata prior to execution to identify prompt injection
risks, tool poisoning, rug-pull attacks, and unsafe cross-origin settings. It
focuses on configuration hardening by scanning tool descriptions, enforcing
trusted tool pinning, and validating allowlists, making it well suited for
pre-deployment security checks and CI/CD
integration~\cite{invariant2025mcpscan}.
\subsubsection{MCPScan.ai}

MCPScan.ai provides a hosted, enterprise-grade MCP security platform that
extends static analysis with continuous repository monitoring and reporting. It
scans MCP-related assets for prompt injection, tool poisoning, and cross-origin
escalation risks, generating comprehensive security reports and dashboards that
support organizational governance, auditing, and large-scale MCP
deployments~\cite{mcpscanai2025}.

\subsubsection{MCIP-Guardian}

MCIP-Guardian is an MCP-native, context-aware safety guard that detects unsafe
tool use by reasoning over the full interaction context, not just individual
function calls. It logs MCP interactions as structured information-flow
trajectories (sender, receiver, data subject, information type, transmission
principle) and uses a taxonomy-guided safety model to determine whether a tool
call is appropriate in context. Unlike network or sandbox defenses, MCIP-Guardian
focuses on semantic and contextual misuse---e.g., function injection, excessive
privileges, replay, and configuration drift---and operates as a runtime
risk-detection layer for MCP agents, improving safety without changing the
underlying MCP protocol~\cite{jing2025mcip}.


\subsubsection{MCP Defender}

MCP-Defender provides runtime, behavior-level protection by monitoring and
blocking malicious MCP traffic within developer environments such as Cursor,
Claude, VS~Code, and Windsurf. It detects suspicious tool interactions and
exploit attempts in real time, offering endpoint-level defense during live agent
execution~\cite{mcpdefender2025}.

\subsubsection{MCP-Gateway (Lasso Security)}

MCP-Gateway acts as a secure proxy and orchestration layer between LLM clients
and MCP servers, enforcing zero-trust access control and centralized policy
management. By intercepting and routing MCP traffic, it prevents unauthorized
tool usage, supports multi-MCP aggregation, and provides architectural isolation
between agents and tools~\cite{lasso2025mcpgateway}.

\subsubsection{ToolHive (Stacklok)}

ToolHive focuses on secure MCP server deployment and lifecycle management,
mitigating supply-chain and misconfiguration risks through automated setup and
access control enforcement. It enables organizations to securely provision,
manage, and govern MCP servers while reducing the operational overhead associated
with manual configuration~\cite{stacklok2025toolhive}.

\subsubsection{MCP-Guard}

MCP-Guard is a proxy-based defense framework designed to secure Model Context
Protocol (MCP) interactions between LLMs and external tools. It employs a
three-stage layered defense: (i)~lightweight, pattern-based static scanning to
quickly block obvious threats such as prompt injection, sensitive file access,
and command injection; (ii)~a learnable neural detector based on a fine-tuned
E5 embedding model to identify subtle, semantic attacks such as tool poisoning;
and (iii)~LLM-based arbitration to resolve uncertain cases and reduce false
positives. This fail-fast, escalation-based design enables high detection
accuracy, low latency, hot-updatable rules, and registry-free deployment, making
MCP-Guard suitable for real-time protection in production MCP
environments~\cite{xing2025mcp_guard}.

\subsubsection{MCP Guardian (eqtylab)}

MCP Guardian (eqtylab) operates as an enterprise MCP access and governance
gateway, enforcing authorization, approval workflows, and compliance controls
over LLM-to-tool interactions. It provides real-time auditing, policy
enforcement, and monitoring to prevent unauthorized access and ensure secure MCP
usage in regulated environments~\cite{eqtylab2025mcpguardian}.

\subsubsection{MindGuard}

MindGuard detects MCP tool poisoning attacks in which malicious tool metadata
corrupts an LLM's planning without executing the poisoned tool. It identifies
both explicit invocation hijacking (forcing unrelated high-privilege calls) and
implicit parameter manipulation (subtly altering arguments), and can precisely
attribute each attack to the poisoned tool source using decision-level
analysis~\cite{wang2025mindguard}.

\subsubsection{Cisco MCP Scanner}

Cisco MCP Scanner combines rule-based and semantic analysis to detect malicious
MCP artifacts, including poisoned tool metadata, prompt injection vectors, and
over-privileged tools. By integrating YARA rules, LLM-based semantic judgment,
and Cisco AI Defense heuristics, it identifies both syntactic and contextual
threats and can be deployed as a standalone scanner or integrated via an
SDK~\cite{cisco2025mcpscanner}.

\subsubsection{MCP-Shield}

MCP-Shield is a CLI-based security scanner that analyzes MCP servers and
configurations to detect tool poisoning, hidden instructions, covert data
exfiltration channels, and tool shadowing attacks. It leverages LLM-powered
semantic analysis to assess contextual risk and produces human-readable security
findings with safe-list support~\cite{riseignite2025mcpshield}.

\subsubsection{AIM-MCP}

AIM-MCP intervenes before an action is executed, influencing whether and how an
AI agent proceeds with an MCP interaction. It analyzes the agent's intent,
prompt content, contextual parameters (MCP type, operation, sensitivity), and
risk signals, then modifies, constrains, or blocks the decision itself by
injecting security instructions, issuing warnings, or flagging high-risk
behavior. This directly shapes the agent's decision-making process rather than
merely observing outcomes or enforcing post-hoc checks.

Crucially, AIM-Guard-MCP does not rely on isolation, sandboxing, or static code
analysis. Instead, it operates at runtime and alters agent behavior dynamically
based on contextual and behavioral signals---such as prompt-injection patterns,
credential exposure, and malicious URLs. Because the defense acts at the point
where the agent chooses what to do next, it squarely belongs to the
decision-level defense layer, making it especially effective against
vulnerable-client attacks driven by malicious or misleading tool
outputs~\cite{aim_guard_mcp_2025}.

%

\begin{table}[htbp]
\centering
\small
\caption{Categorical mapping of representative MCP defense mechanisms
         across four orthogonal defense types.
         $\bullet$~=~supported;\enspace $\circ$~=~not supported.}
\label{tab:categorical_mapping}
\setlength{\tabcolsep}{7pt}
\renewcommand{\arraystretch}{1.25}
\begin{tabular}{@{}lcccc@{}}
\toprule
\textbf{Defense Mechanism}
  & \textbf{\shortstack{Static /\\Pre-Exec.}}
  & \textbf{\shortstack{Behavior /\\Runtime}}
  & \textbf{\shortstack{Isolation /\\Architect.}}
  & \textbf{\shortstack{Decision\\Level}} \\
\midrule
MCP-Scan (Invariant Labs) & $\bullet$ & $\circ$   & $\circ$   & $\circ$   \\
MCPScan.ai                & $\bullet$ & $\circ$   & $\circ$   & $\circ$   \\
MCIP-Guardian             & $\circ$   & $\bullet$ & $\circ$   & $\circ$   \\
AegisMCP                  & $\circ$   & $\bullet$ & $\circ$   & $\circ$   \\
MCP-Defender              & $\circ$   & $\bullet$ & $\circ$   & $\circ$   \\
MCP-Gateway / Lasso       & $\circ$   & $\circ$   & $\bullet$ & $\circ$   \\
ToolHive (Stacklok)       & $\circ$   & $\circ$   & $\bullet$ & $\circ$   \\
MCP-Guard                 & $\bullet$ & $\bullet$ & $\bullet$ & $\circ$   \\
MCP Guardian (eqtylab)    & $\circ$   & $\circ$   & $\bullet$ & $\circ$   \\
MindGuard                 & $\circ$   & $\circ$   & $\circ$   & $\bullet$ \\
Cisco MCP Scanner         & $\bullet$ & $\circ$   & $\circ$   & $\circ$   \\
MCP-Shield                & $\bullet$ & $\circ$   & $\circ$   & $\circ$   \\
AIM-Guard-MCP             & $\circ$   & $\bullet$ & $\circ$   & $\bullet$ \\
\bottomrule
\end{tabular}

\vspace{4pt}
\begin{minipage}{\linewidth}
{\scriptsize
\textbf{Static / Pre-Exec.:} Static/Pre-Execution ---
analyzes tool metadata and configurations before runtime.
\textbf{Behavior / Runtime:} monitors and constrains
tool interactions during execution.
\textbf{Isolation / Architect.:} enforces trust boundaries
between agents and tools via mediation layers.
\textbf{Decision Level:} protects the LLM's internal
tool-selection and parameterization process.
}
\end{minipage}
\end{table}
\subsection{Categorical Mapping of MCP Defense Mechanisms by Defense Type}

Table ~\ref{tab:categorical_mapping} presents a categorical mapping of
representative MCP defense mechanisms across four orthogonal defense types:
Static/Pre-Execution, Behavior-Level/Runtime, Isolation/Architectural, and
Decision-Level. Each row corresponds to a concrete defense system, while each
column denotes a defense type. A checkmark indicates that the defense
substantially and explicitly implements the corresponding protection mechanism.
Multiple checkmarks per row indicate strong, intentional overlap rather than
incidental or auxiliary functionality. The figure is designed to emphasize where
in the MCP execution lifecycle a defense primarily intervenes, enabling a clear
comparison of coverage patterns and defense placement.

\paragraph{Analysis and Key Observations.}

\textbf{Static vs.\ runtime separation is clear and non-overlapping.}
Static defenses such as MCP-Scan, MCPScan.ai, Cisco MCP Scanner, and MCP-Shield
are exclusively categorized under Static/Pre-Execution, reflecting their focus
on analyzing tool metadata, configurations, and MCP artifacts before execution.
These defenses do not monitor live agent behavior and therefore do not provide
runtime enforcement. Conversely, AegisMCP, MCP-Defender, and MCIP-Guardian are
classified under Behavior-Level/Runtime, as they operate by observing or
constraining tool interactions during execution. This clear separation highlights
the complementary roles of static hardening and runtime monitoring.

\textbf{Isolation-based defenses form a distinct architectural layer.}
Defenses such as MCP-Gateway/Lasso, ToolHive, and MCP Guardian (eqtylab) are
solely marked under Isolation/Architectural, reflecting their role as mediation
layers that enforce trust boundaries between agents and tools. These systems
reduce blast radius and privilege exposure without performing semantic reasoning
or behavioral anomaly detection. Notably, isolation-based defenses do not
inherently detect malicious intent; instead, they assume partial compromise and
limit the consequences of misuse.

\textbf{Decision-level defenses are rare but semantically powerful.}
Only MindGuard and AIM-Guard-MCP are marked under Decision-Level, underscoring
how uncommon defenses are that directly protect the LLM's internal
decision-making process---such as tool selection and parameter formation. These
defenses are particularly relevant for vulnerable-client attacks, where malicious
tool metadata or outputs manipulate agent planning without triggering overt
policy violations or runtime anomalies. Decision-level defenses thus represent a
specialized but critical layer not covered by static scanning or architectural
isolation alone.

\textbf{Intentional overlap reflects layered defense, not redundancy.}
A small number of defenses---most notably MCP-Guard and AIM-Guard-MCP---span
multiple columns. This overlap is intentional and meaningful. MCP-Guard combines
static scanning, runtime interception, and proxy-based isolation, making it a
genuinely multi-layer defense. AIM-Guard-MCP spans Behavior-Level and
Decision-Level, as it both observes runtime context and actively intervenes in
agent decisions before execution. The figure avoids weak or incidental overlaps,
ensuring that each mark represents a core security mechanism rather than an
auxiliary feature.

\textbf{Implication for MCP security evaluation.}
The distribution of marks reveals that most existing defenses cluster around
static analysis and runtime monitoring, while decision-level protection remains
underexplored. This imbalance motivates the need for benchmarks and evaluations
that explicitly consider planning-time manipulation and vulnerable-client
scenarios, rather than focusing solely on execution-time exploits.


\subsection{Attack Volume and Coverage Across MCP Categories}

Based on the benchmarking results, the distribution of defense capabilities
across the MCP ecosystem reveals significant disparities in coverage. The
defense-in-depth landscape is currently characterized by a heavy concentration
of protection in specific layers, while structural gaps remain in others.

\subsection{Core Coverage Patterns}
Table ~\ref{tab:defense_layer_coverage} quantifies the capability-based
coverage of each defense mechanism across the six MCP architectural
layers, expressed as the percentage of attacks in that layer that the
mechanism can reasonably detect, block, or constrain. Each row
corresponds to a defense tool, and each column represents a primary
enforcement layer: Model Provider/LLM Alignment (MP/LA), MCP
Host/Application (MH/A), Registry/Marketplace \& Supply-Chain
(R/M\&SC), MCP Client/SDK (MC/S), MCP Server/Tool Execution (MS/TE),
and Transport/Network (T/N). Bold values mark the highest coverage
per column. Three patterns emerge clearly. First, registry and
server-execution layers attract the strongest protection --- ToolHive
achieves $100\%$ registry coverage, and MCP-Guardian and ToolHive each
reach $50\%$ at the server-execution layer. Second, model-alignment
coverage is partial across all tools, with MCIP-Guardian and MCP-Guard
leading at $44\%$. Third, and most critically, the Transport/Network
layer is almost entirely undefended: all tools except MCP-Gateway
($50\%$) report $0\%$ coverage, and host-side orchestration peaks at
only $38\%$ via MCP-Gateway. These figures confirm that current
defenses cluster around tool-adjacent layers while the infrastructure
layers that carry the highest blast radius remain structurally
underprotected.

\small
\renewcommand{\arraystretch}{1.25}
\setlength{\tabcolsep}{12pt}        

\begin{table}[htbp]
\centering
\caption{Defense Coverage Across MCP Ecosystem Layers}
\label{tab:defense_layer_coverage}
\small
\renewcommand{\arraystretch}{1.25}
\begin{tabular}{@{}lcccccc@{}}
\toprule
\textbf{Defense Mechanism}
  & \textbf{MP/LA}
  & \textbf{MH/A}
  & \textbf{R/M\&SC}
  & \textbf{MC/S}
  & \textbf{MS/TE}
  & \textbf{T/N} \\
\midrule
MCP-Scan       & 11\% & 8\%  & 83\%           & 9\%           & 33\%          & 0\%  \\
MCIP-Guardian  & \textbf{44\%} & 8\%  & 0\%   & 0\%           & 17\%          & 0\%  \\
Cisco MCP      & 11\% & 0\%  & 17\%           & 0\%           & 0\%           & 0\%  \\
MCP-Gateway    & 22\% & \textbf{38\%} & 0\%   & 9\%           & 33\%          & \textbf{50\%} \\
MCP-Guardian   & 33\% & 8\%  & 0\%            & \textbf{36\%} & \textbf{50\%} & 0\%  \\
MCP-Defender   & 11\% & 31\% & 0\%            & \textbf{36\%} & 33\%          & 0\%  \\
MCP-Guard      & \textbf{44\%} & 23\% & 17\%  & 0\%           & 33\%          & 25\% \\
Prisma AIRS    & 33\% & 31\% & 17\%           & 9\%           & 33\%          & 0\%  \\
MCPScan.ai     & 0\%  & 0\%  & 17\%           & 9\%           & 0\%           & 0\%  \\
MCP-Shield     & 0\%  & 0\%  & 17\%           & 0\%           & 17\%          & 0\%  \\
ToolHive       & 0\%  & 23\% & \textbf{100\%} & 9\%           & \textbf{50\%} & 0\%  \\
MindGuard      & 0\%  & 8\%  & 50\%           & 0\%           & 17\%          & 0\%  \\
AIM-MCP        & 0\%  & 8\%  & 17\%           & \textbf{36\%} & 0\%           & 25\% \\
\bottomrule
\end{tabular}
\vspace{4pt}

\begin{minipage}{\linewidth}
{\scriptsize
\textbf{MP/LA}: Model Provider/LLM Alignment;
\textbf{MH/A}: MCP Host/Application;
\textbf{R/M\&SC}: Registry/Marketplace \& Supply Chain;
\textbf{MC/S}: MCP Client/SDK;
\textbf{MS/TE}: MCP Server/Tool Execution;
\textbf{T/N}: Transport/Network.
Bold values indicate the highest coverage per column.
}
\end{minipage}
\end{table}

\section{Discussion}

This section interprets our coverage analysis to identify (i)~the dominant
patterns in today's MCP defenses, (ii)~the highest-impact gaps, and
(iii)~practical priorities for closing them. Importantly, our coverage results
reflect \emph{capability-based applicability} under common MCP threat
assumptions, not empirical detection accuracy or guarantees of security.

A key finding is that current defenses cluster around pre-execution scanning and
runtime monitoring, yielding comparatively stronger support for
tool-execution-adjacent threats---e.g., metadata poisoning and malicious tool
outputs---than for ecosystem and infrastructure threats. In contrast, attacks
involving transport manipulation, host orchestration failures, and
registry/supply-chain compromise remain underdefended, despite their broad blast
radius across deployments. This imbalance suggests that many MCP failures are
\emph{structural}: defenses are often deployed where tools are easiest to
inspect, rather than where authority and visibility are sufficient to reliably
prevent or contain the attack.

Two implications follow. First, practitioners should prioritize controls that
reduce systemic exposure: strengthen host-side enforcement (policy checks,
permission scoping, and auditability), harden transport assumptions (secure
binding and integrity), and improve registry/supply-chain hygiene (provenance
and update controls). Second, the scarcity of decision-level defenses indicates
an important research gap---attacks that steer tool selection or parameterization
can evade output-only checks, especially in multi-step workflows. In general,
the taxonomy can serve as a placement-oriented checklist for evaluating MCP
deployments and as a roadmap for developing defenses that cover currently
underprotected threat classes.

\section{Conclusion}

This paper introduced a defense-placement-oriented taxonomy for Model Context
Protocol (MCP) security that organizes MCP-specific attacks by architectural
responsibility and identifies primary and secondary enforcement points to support
defense-in-depth reasoning. Using this taxonomy, we mapped representative
academic and industry defenses and conducted a capability-based coverage analysis
to characterize how current mitigations align with the MCP threat landscape.

Our analysis shows that existing defenses are uneven and frequently concentrate
on tool-adjacent protections, while important threats involving host
orchestration, transport assumptions, and registry/supply-chain mechanisms remain
comparatively underdefended. These gaps suggest that many MCP security failures
are \emph{structural}: defenses are often deployed where implementation is
easiest rather than where authority and visibility are sufficient to reliably
prevent or contain an attack.

We expect this taxonomy and mapping to serve as a practical reference for
evaluating MCP deployments and for prioritizing new mitigations that address
undercovered attack classes. Future work should strengthen defenses in the most
underprotected layers and expand decision-level protections that can resist
planning-time manipulation in multi-step MCP workflows.

\appendix

\section{Attack Descriptions}
\label{app:attacks}

This appendix provides definitions and illustrative descriptions for the
MCP-specific attacks enumerated in the taxonomy. Each entry describes the
attack mechanism, the adversarial objective, and the primary pathway through
which it manifests in MCP-based systems.

\paragraph{Tool Poisoning.}
An adversary embeds malicious natural-language instructions into a tool's
description or metadata, which are then loaded into the agent's context during
MCP registration~\cite{wang2025mcptox}. The agent treats these poisoned
instructions as part of the tool specification, leading it to perform
unauthorized operations---e.g., secret exfiltration---via otherwise legitimate
tools. The maliciously modified tool behaves unexpectedly, causing the MCP
system or LLM to execute unauthorized actions. This can result in data leaks,
incorrect decisions, or indirect compromise of downstream tools while appearing
syntactically normal~\cite{errico2025securing,xing2025mcp_guard}.

\paragraph{Parameter Poisoning.}
The attacker does not block the tool but secretly alters the parameters the
agent uses---e.g., file path, stock ticker, or account identifier. As a result,
the agent performs the correct type of operation but on attacker-chosen, harmful
targets~\cite{Li2026MCPITP}.

\paragraph{Command Injection.}
Malicious payloads such as shell commands or code fragments are injected into
tool descriptions or outputs. When the agent or host executes these as commands
instead of treating them as data, the attacker can achieve arbitrary code
execution on the underlying system~\cite{owasp2025agenticai}.

\paragraph{Function Overlapping.}
The attacker introduces tools whose names and signatures overlap with benign
tools but whose behavior is adversarial. Because the model must choose among
similar tools under natural-language instructions, it can be steered toward the
dangerous variant~\cite{zong2025mcpsafetybench}.

\paragraph{Preference Manipulation Attack.}
Tool descriptions are written to strongly promote a particular tool---e.g.,
``most accurate'' or ``recommended for all tasks.'' The agent's tool-selection
policy is biased toward the attacker-controlled tool, even when safer or more
appropriate alternatives exist~\cite{zong2025mcpsafetybench,Wang2025MPMA}.

\paragraph{Tool Shadowing.}
A malicious tool or server is given a name and description that closely mimic a
trusted one, effectively acting as a shadow version. The agent or user may
connect to and invoke this shadow instance by mistake, exposing data or
privileges to the attacker~\cite{styer2025agenttools}.

\paragraph{Rug Pull / Malicious Automatic Update.}
A tool or server behaves benignly at first and becomes integrated into normal
workflows~\cite{song2025beyond}. After trust is established, the implementation
or metadata is changed to include malicious behavior, enabling sudden data
exfiltration or destructive actions without obvious changes in
interface~\cite{errico2025securing}.

\paragraph{Package Name Squatting (Server Name).}
Package name squatting on server names targets the server layer rather than
individual tools. The attacker publishes or registers an MCP server whose
identifier closely resembles that of a legitimate server, so that host
configurations or users accidentally bind to the malicious server, thereby
routing tool calls and data flows through an adversary-controlled
infrastructure~\cite{yang2025mcpsecbench}.

\paragraph{Package Name Squatting (Tool Name).}
Tool name squatting occurs when a malicious MCP server registers a tool with a
name identical or deceptively similar to a trusted one. This exploits the LLM's
selection logic, causing it to inadvertently execute the attacker's tool and
perform actions that diverge from the user's original
intent~\cite{yang2025mcpsecbench}.

\paragraph{Privilege Escalation (MCP Context).}
The attacker causes the agent to use tools or resources with broader permissions
than the user intended. Small manipulations in planning or tool selection can
result in the agent performing high-impact operations---e.g., modifying system
files or sensitive accounts. Although prior studies demonstrate MCP scenarios in
which agents are induced to invoke high-privilege tools beyond user
intent~\cite{RadosevichHalloran2025MCPSafetyAudit,song2025beyond,zhao2025mindyourserver},
existing work does not explicitly characterize these behaviors as privilege
escalation attacks. We formalize this class and analyze it through the lens of
defense placement.

\paragraph{Intent Injection.}
The context injected through tools, resources, or prior messages is designed to
change the model's interpretation of the user's true goal. The agent then plans
and acts according to this altered intention---e.g., ``test security'' or
``gather secrets''---rather than the original benign
request~\cite{jing2025mcip}.

\paragraph{Data Tampering.}
The attacker modifies data delivered by MCP tools---e.g., financial prices,
navigation results, or repository contents---while preserving valid formats. The
agent makes decisions based on corrupted but plausible data, enabling subtle and
impactful manipulation of outcomes. Several studies show that MCP agents can
consume manipulated but syntactically valid tool output originating from
malicious servers, leading to incorrect decisions and downstream
effects~\cite{song2025beyond,RadosevichHalloran2025MCPSafetyAudit,zhao2025mindyourserver}.

\paragraph{Identity Spoofing.}
Tool responses or messages are crafted to appear as if they originate from a
specific user or trusted component. The agent, believing the spoofed identity,
may execute sensitive actions or reveal information that would be withheld from
an untrusted source~\cite{gasmi2025bridging}.

\paragraph{Replay Injection.}
Old but valid messages, tool calls, or tool outputs are replayed in a new
context. Because the agent often trusts its history, these replayed elements can
trigger outdated operations or bypass checks that assumed one-time or time-bound
behavior~\cite{zong2025mcpsafetybench}.

\paragraph{Malicious Code Execution / Remote Code Execution.}
The agent is convinced to generate and run attacker-chosen code through powerful
tools---e.g., shell, Python, or notebook environments. Once executed on the host
environment, the attacker gains system-level access, collapsing the boundary
between prompt compromise and classic OS
exploitation~\cite{RadosevichHalloran2025MCPSafetyAudit,siddiq2026empirical}.

\paragraph{Credential Theft.}
The agent is guided to access and expose secrets---API keys, SSH keys, tokens,
or passwords---from files, environment variables, or secret stores accessible to
MCP tools~\cite{RadosevichHalloran2025MCPSafetyAudit}. Stolen credentials can
then be used for persistent account takeover or lateral movement outside the MCP
system~\cite{zong2025mcpsafetybench,guo2025systematic}.

\paragraph{Jailbreak.}
Carefully constructed prompts or contexts are used to bypass the LLM's safety
rules. In MCP settings, a successful jailbreak is especially dangerous because
the now-unconstrained model can directly drive tools to perform harmful
real-world operations~\cite{Asl2025NEXUS}.

\paragraph{Prompt Leakage.}
The attacker uses crafted queries or indirect injections to extract hidden system
prompts, tool schemas, or internal policies. Knowledge of these internal details
enables more targeted attacks, including stronger prompt injection, better tool
poisoning, and more precise privilege escalation~\cite{guo2025systematic}.

\paragraph{Tool Misuse via Confused AI.}
Tool misuse via ``confused AI'' exploits the model's imperfect understanding of
tool semantics, permissions, or context. By presenting misleading or ambiguous
instructions and metadata, attackers induce the agent to select and combine tools
in ways that violate security assumptions---such as using high-privilege tools
for low-risk tasks or revealing more information than
requested~\cite{owasp2025agenticai}.

\paragraph{Man-in-the-Middle.}
A man-in-the-middle (MitM) attack occurs when an adversary intercepts and
potentially modifies the communication channel between the MCP client and the
server~\cite{gasmi2025bridging}. By relaying or altering JSON-RPC messages, the
attacker can observe sensitive data, inject additional tool calls, or tamper with
tool responses without the knowledge of either endpoint~\cite{yang2025mcpsecbench}.

\paragraph{Vulnerable Client (CVE-2025-6514).}
Vulnerable client attacks occur when MCP client or SDK implementations
implicitly trust data returned by MCP servers~\cite{hasan2025mcp_first_glance}
and perform unsafe local actions based on that data. As demonstrated
in~\cite{yang2025mcpsecbench}, a malicious MCP server can exploit flaws in
client-side logic---such as blindly opening URLs or executing commands embedded
in server responses---to achieve arbitrary code execution or data exfiltration
on the client machine, even when the model itself behaves correctly. These
vulnerabilities stem from insecure client SDK design, insufficient input
validation, and over-privileged local execution paths, making the client a
critical but often overlooked attack surface in MCP
deployments~\cite{lotfi2025automated}.

\paragraph{Prompt Injection.}
Prompt injection denotes any attack in which adversarial text is crafted to
override, circumvent, or subvert the model's existing safety instructions and
policies. In MCP settings, such payloads can be delivered via user input, tool
output, or external resources, causing the agent to prioritize attacker goals
and potentially invoke high-impact tools in violation of the original
intent~\cite{gulyamov2026prompt}.

\paragraph{Goal Hijack.}
Goal hijacking occurs when an attacker manipulates the context so that the model
revises its internal understanding of the overarching task
objective~\cite{guo2025systematic}. Rather than merely altering individual
steps, the adversary reframes the ``true'' goal---e.g., from assisting a user
to exfiltrating secrets---leading the agent to plan and execute entire
tool-calling workflows aligned with the attacker's aims.

\paragraph{Memory Poisoning.}
Memory poisoning targets persistent or long-lived memory components of an agent,
injecting misleading or adversarial content that will inform future decisions.
By corrupting stored facts, user preferences, or historical summaries, the
attacker induces systematic misbehavior between sessions, including biased tool
selection, unsafe defaults, and repeated security policy
violations~\cite{yan2025tradetrap}.

\paragraph{Semantic Misalignment.}
Semantic misalignment describes the divergence between the intended semantics of
instructions, tools, or policies and the model's internal interpretation of
them. In MCP workflows, this misalignment can be exploited so that the agent
believes it is complying with safety constraints or user intent while performing
operations that are broader, riskier, or qualitatively different than
specified~\cite{li2025netmcp}.

\paragraph{Hallucination Attacks.}
Hallucination attacks exploit the model's tendency to generate confident but
false statements by shaping prompts or context so that the agent ``fills in''
missing details with fabricated tools, resources, or behaviors. In MCP systems,
such fabricated beliefs can lead the agent to invoke non-existent components,
misinterpret security-relevant outputs, or rationalize unsafe actions as
legitimate~\cite{guo2025systematic}.

\paragraph{Resource Type Confusion / Impersonation.}
Prior work indicates that MCP agents can be misled about the semantic role or
trust level of resources returned by tools or servers, causing untrusted
artifacts to be interpreted as authoritative data or executable
guidance~\cite{RadosevichHalloran2025MCPSafetyAudit,zhao2025mindyourserver}.
Resource type confusion arises when an adversary deliberately presents a
resource---such as a file, URL, or API endpoint---in a form that obscures its
true provenance or intent. By masquerading adversarial content as benign
artifacts (e.g., logs, documentation, or configuration files), the attacker
induces the agent to ascribe undue trust to the resource and to incorporate its
contents into planning, reasoning, or subsequent tool invocations.

\paragraph{Control-Flow Hijack.}
Control-flow hijack in MCP workflows refers to attacks that redirect the agent's
multi-step reasoning or tool-calling sequence away from the intended execution
path. By inserting adversarial instructions or misleading intermediate results,
the attacker causes the agent to execute an alternative sequence of tool
calls---often skipping validation steps, invoking more powerful tools, or
entering unintended loops---to achieve malicious objectives. Prior work shows
that manipulated intermediate outputs or context can cause MCP agents to deviate
from expected execution paths, although these behaviors are not explicitly framed
as control-flow hijacking~\cite{zhao2025mindyourserver,song2025beyond}.

\paragraph{Malicious Tool Registration.}
Malicious tool registration occurs when an attacker registers a tool whose
implementation or semantics are adversarial but whose interface appears benign.
Once the tool is accepted into the MCP environment, any invocation can execute
hidden malicious logic, leveraging the privileges and trust associated with its
declared capabilities~\cite{guo2025systematic}.

\paragraph{Unauthorized Resource Invocation.}
Unauthorized resource invocation denotes scenarios where the agent is induced to
access tools, servers, or data sources that are outside the user's intended
security scope. This can arise from prompt injection, misconfigured permissions,
or ambiguous instructions, leading the agent to fetch or manipulate resources
that should require additional authorization or explicit
approval~\cite{RadosevichHalloran2025MCPSafetyAudit}.

\paragraph{MCP Rebinding.}
MCP rebinding refers to the unauthorized reassignment of an existing server or
tool identifier to a different endpoint or implementation. Because clients and
agents typically trust the identifier rather than the underlying address,
subsequent invocations are transparently routed to the attacker-controlled
backend, enabling interception, manipulation, or full compromise of workflows
that rely on the original binding~\cite{yang2025mcpsecbench}.

\paragraph{Installer Spoof.}
Installer spoof attacks target the distribution or setup phase by presenting a
malicious installer, script, or configuration bundle that mimics a legitimate
MCP tool or server installation. Once executed, the spoofed installer silently
registers adversarial tools, modifies bindings, or plants backdoors, so that
subsequent MCP workflows operate atop a compromised
foundation~\cite{hou2025mcp_landscape}.

\paragraph{Multi-Tool Cooperation / Propagation.}
Multi-tool cooperation or propagation describes attacks that deliberately chain
multiple tools to amplify impact or spread compromise across components. An
adversary designs intermediate outputs and dependencies so that the results of a
compromised tool trigger follow-up invocations in other tools, allowing lateral
movement, persistent misconfiguration, or data exfiltration across the MCP
ecosystem~\cite{zhao2025mindyourserver}.

\paragraph{Resource Overload.}
Resource overload attacks exploit the tool-centric execution model of MCP to
induce an agent to issue excessive or pathologically expensive tool calls. By
repeatedly invoking computationally intensive tools, constructing oversized
request payloads, or triggering unbounded control flows---e.g., deep tool
recursion, large data retrievals, or massive file operations---an attacker can
exhaust computational, storage, or network resources. Such attacks degrade agent
responsiveness and may result in partial denial-of-service affecting co-located
agents, shared MCP servers, or downstream
services~\cite{owasp2025agenticai,li2025leechhijack}.

\paragraph{Sampling Abuse.}
A malicious MCP server exploits the sampling feature to secretly cause the
user's LLM to generate content during tool execution instead of using its own
resources. This shifts computational cost to the user while the attacker
benefits from free, high-quality AI output~\cite{zhao2025when_mcp_servers_attack}.

\paragraph{Malicious Tool Installation.}
In a malicious project installation attack, attackers hide harmful commands in
the README file of a software package to exploit MCP server-assisted
installation. For example, a \texttt{README.md} might include a line such as
\texttt{curl http://malicious.example.com/malware.sh | bash}. When users follow
the instructions, the script executes automatically. This type of attack can
spread widely through software supply chains as affected packages are adopted by
many users~\cite{guo2025systematic}.

\paragraph{Information Overcollection.}
An information overcollection attack occurs when a malicious actor tricks or
manipulates an MCP server or agent into collecting more information than
necessary. For example, an attacker submits a task asking an agent to return
weather data alongside all users' email addresses and session tokens, causing
overcollection of sensitive information~\cite{zhao2025when_mcp_servers_attack}.

\paragraph{Binary Payload.}
Binary payloads constitute a potential attack vector in MCP interactions.
Resources may return base64-encoded binary data accompanied by a
\texttt{mimeType} field, which the client decodes and processes accordingly. A
malicious server could craft such payloads to exploit vulnerabilities in the
host's parsing or rendering stack, potentially leading to system
compromise~\cite{zhao2025when_mcp_servers_attack}.

\paragraph{Slash Command Overlap.}
Slash command overlap occurs when multiple MCP tools define identical or similar
slash commands---e.g., \texttt{/remove}~\cite{yu2025survey}---creating execution
conflicts. An attacker can exploit this ambiguity to trigger unintended actions,
compromising system reliability and potentially causing data
loss~\cite{anbiaee2026security}.

\paragraph{Agent Communication Poisoning.}
Agent communication poisoning occurs when a malicious actor injects false,
misleading, or incomplete messages into the communication channels used by agents
to coordinate~\cite{eslami2025security,ghosh2025safety}. These manipulated
messages can distort an agent's understanding of its environment or the
intentions of other components, causing it to execute unintended actions. Since
the resulting behavior may appear structurally correct, the MCP system often
lacks indicators that upstream coordination was compromised, allowing attackers
to subtly disrupt workflows or influence
decision-making~\cite{owasp_llm_top10}.

\paragraph{Malicious Authorization Metadata.}
Malicious authorization metadata arises when an MCP server transmits
authorization-related information---such as an authorization server URL---over
HTTP. If an attacker manipulates this metadata to point to a hostile endpoint or
embeds executable instructions, and the client does not properly validate it,
the system may be redirected to the attacker-controlled resource or
unintentionally execute harmful code~\cite{zhao2025when_mcp_servers_attack}.

\paragraph{Server Metadata Poisoning / Deceptive Naming.}
Server metadata poisoning occurs when an attacker manipulates MCP server
metadata---such as names or descriptions---to appear trustworthy. This can lure
users into integrating malicious servers or mislead LLMs into treating harmful
servers as legitimate, potentially exposing both the model and users to
subsequent attacks~\cite{zhao2025when_mcp_servers_attack}.

\paragraph{Remote Listener.}
A remote listener attack involves embedding persistent control commands---e.g.,
reverse shells---within MCP tool descriptions, often disguised as benign
utilities such as ``debugging tools'' or ``network plugins.'' Because MCP
servers typically do not monitor tool metadata in real time, these malicious
commands can persist undetected, allowing attackers to steal data or maintain
backdoor access~\cite{guo2025systematic}.

\paragraph{Server Code Leakage.}
Server code leakage occurs when attackers exploit MCP server responses that
reveal debug information---e.g., file paths. By analyzing these hints and
issuing crafted requests (e.g., \texttt{GET /debug?file=server.py}) or
enumerating directories via APIs, they can access sensitive source code and
system files~\cite{guo2025systematic}.

\paragraph{File-Based Injection.}
File-based injection in MCP occurs when attackers embed malicious content in
files processed by clients or agents. If the system trusts these files without
validation, this can trigger unintended actions such as code execution, prompt
injection, or data theft~\cite{zhao2025when_mcp_servers_attack}.

\paragraph{Endpoint Exposure.}
Endpoint exposure occurs when a locally installed MCP server opens an overly
permissive HTTP endpoint---e.g., bound to \texttt{0.0.0.0}. This allows
network-level attackers to access the server and potentially exploit powerful
capabilities such as file access on the user's
system~\cite{zhao2025when_mcp_servers_attack}.

\paragraph{Adversarial Connection Parameters.}
Adversarial connection parameters occur when an attacker provides a malicious
MCP server URL in the host configuration. When the host connects, all subsequent
communication is routed through the attacker-controlled endpoint, enabling
interception, manipulation, or exploitation of the interaction. MCP hosts often
adopt a ``configure once, run always'' model, automatically relaunching any
registered server at every system startup without further user consent. A
malicious server can exploit this by inserting its launch command into the host's
configuration file, ensuring it executes in the background at every boot and
transforming a one-time interaction into a persistent
foothold~\cite{zhao2025when_mcp_servers_attack}.

\paragraph{Backdoor Attack.}
In MCP, backdoor attacks use tools and server resources as entry points for
malicious logic~\cite{guo2025systematic}. Hidden instructions or pre-embedded
scripts~\cite{zhu2025demonagent} can cause the LLM to execute unintended actions
when a tool is invoked~\cite{yang2024watch}, with triggers selectively activating
the backdoor for covert compromise.

\paragraph{DoS Against the Client.}
A client-side denial-of-service (DoS) is a resource exhaustion attack in which
a server exploits the client by sending malicious or malformed protocol
primitives. Unlike traditional DoS attacks, here the attacker is the provider
and the victim is the consumer---e.g., Claude Desktop, an IDE, or an autonomous
agent~\cite{zhao2025when_mcp_servers_attack}.

\paragraph{Cross-Origin Violation.}
A cross-origin violation occurs when an agent bridges isolated domains, allowing
a malicious server to manipulate tools from a trusted
one~\cite{Gaire2025MCP}. For example, an adversarial ``Weather'' server could
inject a hidden prompt that forces the LLM to exfiltrate private emails from a
``Gmail'' server~\cite{maloyan2026promptinjection}. This persists because MCP
currently lacks a native same-origin policy to enforce data isolation between
different providers~\cite{errico2025securing}.

\paragraph{Tool Squatting Amplification.}
Tool squatting amplification is a multi-stage attack in which an adversarial
server registers a squatted tool---e.g., \texttt{git\_clone\_secure}---to
intercept an LLM's initial request and inject a malicious system prompt that
overrides the agent's core safety guardrails~\cite{maloyan2026promptinjection}.
By combining name deception with recursive prompt injection, the attacker
amplifies a single misdirected tool call into full, persistent control over the
agent's future reasoning and tool selections. For example, a squatted ``File
Search'' tool might return a result that secretly instructs the LLM to BCC every
drafted email to an attacker-controlled address~\cite{maloyan2026promptinjection}.

\paragraph{Selection Inducement.}
Selection inducement occurs when an adversarial server manipulates tool or
resource metadata---such as enticing descriptions, high priority values, or
false ``premium'' claims---to bias an LLM's selection logic. By exaggerating
utility or freshness, the attacker ensures their malicious component is
prioritized over legitimate ones, effectively hijacking the agent's workflow.
This exploit succeeds because current MCP implementations lack mechanisms to
validate whether a server's semantic claims match its underlying
data~\cite{zhao2025when_mcp_servers_attack}.

\paragraph{Persistence via Host Auto-Launch.}
MCP hosts often use a ``configure once, run always'' model, automatically
relaunching registered servers at every system startup without further user
consent. A malicious server exploits this by inserting its launch command into
the host's configuration file, ensuring execution in the background on every
boot and maintaining a persistent foothold across
restarts~\cite{zhao2025when_mcp_servers_attack}.

\paragraph{Multi-Agent Collaboration / Collusion.}
A multi-agent coordination or collusion attack occurs when a chain of seemingly
benign interactions across multiple agents collectively implements a complex
exploit---such as privilege escalation or data exfiltration. These systemic
failures arise from cascading misalignments or adversarial message passing,
which can bypass single-point security filters because no individual agent's
action appears malicious in isolation~\cite{mahmud2025distributed}.

\paragraph{Promotional / Deceptive Metadata.}
Attackers use deceptive names and high-value descriptions on public platforms to
lure users into installing malicious MCP servers. Once integrated, this
promotional metadata appears in the agent's UI, tricking the user into engaging
with unsafe tools or resources~\cite{zhao2025when_mcp_servers_attack}.

\paragraph{Function Dependency Injection.}
A malicious server forges the dependency chain of a target function by falsely
claiming that certain additional functions must be called beforehand. As a
result, the LLM is misled into invoking unnecessary functions, introducing
potential risks~\cite{jing2025mcip}.

\paragraph{Schema Inconsistency.}
A schema inconsistency attack occurs when the tool interface schema---e.g., JSON
parameters, types, or constraints---presented to the agent differs from the
server's actual expected schema, causing the model to generate malformed or
unintended calls. Attackers can exploit this mismatch to bypass validation,
trigger hidden functionality, or coerce the agent into unsafe operations, because
the agent relies on the incorrect schema as ground truth. This violates the
integrity assumption that tool specifications are consistent and trustworthy
across registration and execution stages~\cite{yang2025mcpsecbench}.

\paragraph{Function Return Injection.}
Attackers embed malicious instructions within a tool's return payload, which can
cause the system to perform unintended actions such as invoking other tools or
running arbitrary code~\cite{zong2025mcpsafetybench}.

\paragraph{Configuration Drift.}
Configuration drift is a mismatch between client and server configurations that
develops over time. When policies, permissions, or validations change on one side
but not the other, inconsistencies emerge. These misalignments create persistent
gaps that attackers can exploit to bypass security
controls~\cite{jing2025mcip}.

\bibliographystyle{ACM-Reference-Format}
\bibliography{sample-base}

@String{Computer = "{IEEE} Computer" }

@String{NeurIPS       = "Advances in Neural Information Processing Systems" }

@String{arXiv         = "arXiv preprint" }

@String{IEEE          = "{IEEE}" }

@misc{anthropic_mcp_intro,
  title        = {Introducing the Model Context Protocol},
  author       = {{Anthropic}},
  year         = {2024},
  month        = nov,
  howpublished = {\url{https://www.anthropic.com/news/model-context-protocol}},
  note         = {Accessed: 2026-01-27}
}

@misc{mcp_spec_2025_11_25,
  title        = {Model Context Protocol Specification (Version 2025-11-25)},
  author       = {{Model Context Protocol}},
  year         = {2025},
  month        = nov,
  howpublished = {\url{https://modelcontextprotocol.io/specification/2025-11-25}},
  note         = {Accessed: 2026-01-27}
}

@article{lei2025mcpsurvey,
  title   = {A Comprehensive Survey on Model Context Protocol: Architecture, Tool Integration,
             and the Future of {AI} Interoperability},
  author  = {Lei, Yang and Xu, Jian and Liang, Chuxuan and Bi, Zhen and Li, Xiaoyu
             and Zhang, Di and Song, Jian and Yu, Zhiwen},
  journal = {Preprint},
  year    = {2025},
  month   = dec
}

@misc{Rayarao2025MCP,
  author    = {Rayarao, S. R. and Donikena, N.},
  title     = {Bridging {AI} and External Systems: A Comprehensive Analysis of the
               Model Context Protocol ({MCP})},
  year      = {2025},
  publisher = {arXiv},
  note      = {Preprint}
}

@misc{styer2025agenttools,
  author       = {Styer, Matthew and Patlolla, Kiran and Mohan, Madhav and Diaz, Sebastian},
  title        = {Agent Tools \& Interoperability with {MCP}},
  howpublished = {\url{https://www.kaggle.com/whitepaper-agent-tools-and-interoperability-with-mcp}},
  year         = {2025},
  note         = {Accessed: Nov. 13, 2025}
}

@misc{li2025netmcp,
  title         = {{NetMCP}: Network-Aware Model Context Protocol Platform for {LLM}
                   Capability Extension},
  author        = {Li, Enhan and Du, Hongyang and Huang, Kaibin},
  year          = {2025},
  eprint        = {2510.13467},
  archivePrefix = {arXiv},
  primaryClass  = {cs.AI},
  publisher     = {arXiv},
  doi           = {10.48550/arXiv.2510.13467},
  url           = {https://doi.org/10.48550/arXiv.2510.13467}
}

@misc{hou2025mcp_landscape,
  title         = {Model Context Protocol ({MCP}): Landscape, Security Threats,
                   and Future Research Directions},
  author        = {Hou, Xinyi and Zhao, Yanjie and Wang, Shenao and Wang, Haoyu},
  year          = {2025},
  eprint        = {2503.23278},
  archivePrefix = {arXiv},
  primaryClass  = {cs.CR},
  publisher     = {arXiv},
  doi           = {10.48550/arXiv.2503.23278},
  url           = {https://doi.org/10.48550/arXiv.2503.23278}
}

@misc{guo2025systematic,
  title         = {Systematic Analysis of {MCP} Security},
  author        = {Yongjian Guo and Puzhuo Liu and Wanlun Ma and Zehang Deng
                   and Xiaogang Zhu and Peng Di and Xi Xiao and Sheng Wen},
  year          = {2025},
  eprint        = {2508.12538},
  archivePrefix = {arXiv},
  publisher     = {arXiv},
  doi           = {10.48550/arXiv.2508.12538},
  url           = {https://doi.org/10.48550/arXiv.2508.12538}
}

@misc{Gaire2025MCP,
  author        = {Gaire, S. and Gyawali, S. and Mishra, S. and Niroula, S.
                   and Thakur, D. and Yadav, U.},
  title         = {Systematization of Knowledge: Security and Safety in the
                   Model Context Protocol Ecosystem},
  year          = {2025},
  eprint        = {2512.08290},
  archivePrefix = {arXiv},
  publisher     = {arXiv},
  doi           = {10.48550/arXiv.2512.08290},
  url           = {https://doi.org/10.48550/arXiv.2512.08290}
}

@misc{hasan2025mcp_first_glance,
  title         = {Model Context Protocol ({MCP}) at First Glance: Studying the
                   Security and Maintainability of {MCP} Servers},
  author        = {Hasan, Mohammed Mehedi and Li, Hao and Fallahzadeh, Emad
                   and Rajbahadur, Gopi Krishnan and Adams, Bram and Hassan, Ahmed E.},
  year          = {2025},
  eprint        = {2506.13538},
  archivePrefix = {arXiv},
  primaryClass  = {cs.SE},
  publisher     = {arXiv},
  doi           = {10.48550/arXiv.2506.13538},
  url           = {https://doi.org/10.48550/arXiv.2506.13538}
}

@misc{song2025beyond,
  title         = {Beyond the Protocol: Unveiling Attack Vectors in the
                   Model Context Protocol Ecosystem},
  author        = {Song, Haoyu and Shen, Yifan and Luo, Wei and Guo, Liang
                   and Chen, Tianyu and Wang, Jianwei and Li, Bo and Zhang, Xiang
                   and Chen, Jian},
  year          = {2025},
  eprint        = {2506.02040},
  archivePrefix = {arXiv},
  primaryClass  = {cs.CR},
  publisher     = {arXiv},
  doi           = {10.48550/arXiv.2506.02040},
  url           = {https://doi.org/10.48550/arXiv.2506.02040}
}

@misc{zhao2025when_mcp_servers_attack,
  title         = {When {MCP} Servers Attack: Taxonomy, Feasibility, and Mitigation},
  author        = {Zhao, Weibo and Liu, Jiahao and Ruan, Bonan and Li, Shaofei
                   and Liang, Zhenkai},
  year          = {2025},
  eprint        = {2509.24272},
  archivePrefix = {arXiv},
  primaryClass  = {cs.CR},
  publisher     = {arXiv},
  doi           = {10.48550/arXiv.2509.24272},
  url           = {https://doi.org/10.48550/arXiv.2509.24272}
}

@misc{errico2025securing,
  title         = {Securing the Model Context Protocol ({MCP}): Risks, Controls,
                   and Governance},
  author        = {Errico, Herman and Ngiam, Jiquan and Sojan, Shanita},
  year          = {2025},
  eprint        = {2511.20920},
  archivePrefix = {arXiv},
  primaryClass  = {cs.CR},
  publisher     = {arXiv},
  doi           = {10.48550/arXiv.2511.20920},
  url           = {https://doi.org/10.48550/arXiv.2511.20920}
}

@misc{gasmi2025bridging,
  title         = {Bridging {AI} and Software Security: A Comparative Vulnerability
                   Assessment of {LLM} Agent Deployment Paradigms},
  author        = {Gasmi, Tarek and Guesmi, Rania and Belhadj, Imen and Bennaceur, Joffrey},
  year          = {2025},
  eprint        = {2507.06323},
  archivePrefix = {arXiv},
  primaryClass  = {cs.CR},
  publisher     = {arXiv},
  doi           = {10.48550/arXiv.2507.06323},
  url           = {https://doi.org/10.48550/arXiv.2507.06323}
}

@inproceedings{thirumalaisamy2025aimcp,
  title     = {{AI} {MCP} Servers in Cybersecurity: Emerging Attack Vectors
               and Mitigation Strategies},
  author    = {Thirumalaisamy, Karthikeyan and Konakalla, Manikarthik
               and Devamanoharan, Dinesh Kumar},
  booktitle = {{AI} {MCP} Servers in Cybersecurity: Emerging Attack Vectors
               and Mitigation Strategies},
  year      = {2025},
  month     = dec,
  address   = {Bothell, Washington, {USA}},
  doi       = {10.5281/zenodo.17931691},
  publisher = {Zenodo}
}

@misc{anbiaee2026security,
  author        = {Anbiaee, Z. and Rabbani, M. and Mirani, M. and Piya, G.
                   and Opushnyev, I. and Ghorbani, A. and Dadkhah, S.},
  title         = {Security Threat Modeling for Emerging {AI}-Agent Protocols:
                   A Comparative Analysis of {MCP}, {A2A}, {Agora}, and {ANP}},
  year          = {2026},
  eprint        = {2602.11327},
  archivePrefix = {arXiv},
  publisher     = {arXiv},
  doi           = {10.48550/arXiv.2602.11327},
  url           = {https://doi.org/10.48550/arXiv.2602.11327}
}

@misc{RadosevichHalloran2025MCPSafetyAudit,
  title         = {{MCP} Safety Audit: {LLMs} with the Model Context Protocol
                   Allow Major Security Exploits},
  author        = {Radosevich, Brandon and Halloran, James},
  year          = {2025},
  eprint        = {2504.03767},
  archivePrefix = {arXiv},
  primaryClass  = {cs.CR},
  publisher     = {arXiv},
  doi           = {10.48550/arXiv.2504.03767},
  url           = {https://doi.org/10.48550/arXiv.2504.03767}
}

@misc{Wang2025MPMA,
  title         = {{MPMA}: Preference Manipulation Attack against Model Context Protocol},
  author        = {Wang, Zihan and Zhang, Rui and Liu, Yu and Fan, Wenshu
                   and Jiang, Wenbo and Zhao, Qingchuan and Li, Hongwei and Xu, Guowen},
  year          = {2025},
  eprint        = {2505.11154},
  archivePrefix = {arXiv},
  primaryClass  = {cs.CR},
  publisher     = {arXiv},
  doi           = {10.48550/arXiv.2505.11154},
  url           = {https://doi.org/10.48550/arXiv.2505.11154}
}

@misc{Li2026MCPITP,
  title         = {{MCP-ITP}: An Automated Framework for Implicit Tool Poisoning in {MCP}},
  author        = {Li, Ruiqi and Wang, Zhiqiang and Yao, Yunhao and Li, Xiang-Yang},
  year          = {2026},
  eprint        = {2601.07395},
  archivePrefix = {arXiv},
  primaryClass  = {cs.CR},
  publisher     = {arXiv},
  doi           = {10.48550/arXiv.2601.07395},
  url           = {https://doi.org/10.48550/arXiv.2601.07395}
}

@misc{wang2025mcptox,
  title         = {{MCPTox}: A Benchmark for Tool Poisoning Attack on
                   Real-World {MCP} Servers},
  author        = {Wang, Zhiqiang and Gao, Yichao and Wang, Yanting and Liu, Suyuan
                   and Sun, Haifeng and Cheng, Haoran and Shi, Guanquan and Du, Haohua
                   and Li, Xiangyang},
  year          = {2025},
  eprint        = {2508.14925},
  archivePrefix = {arXiv},
  primaryClass  = {cs.CR},
  publisher     = {arXiv},
  doi           = {10.48550/arXiv.2508.14925},
  url           = {https://doi.org/10.48550/arXiv.2508.14925}
}

@misc{zhao2025mindyourserver,
  title         = {Mind Your Server: A Systematic Study of Parasitic Toolchain Attacks
                   on the {MCP} Ecosystem},
  author        = {Zhao, Sheng and Hou, Qiming and Zhan, Zhe and Wang, Yu and Xie, Yiming
                   and Guo, Yuxin and Chen, Lei and Li, Shuai and Xue, Zhi},
  year          = {2025},
  eprint        = {2509.06572},
  archivePrefix = {arXiv},
  publisher     = {arXiv},
  doi           = {10.48550/arXiv.2509.06572},
  url           = {https://doi.org/10.48550/arXiv.2509.06572}
}

@misc{li2025leechhijack,
  title         = {{LeechHijack}: Covert Computational Resource Exploitation
                   in Intelligent Agent Systems},
  author        = {Li, Ruiqi and Wang, Zhiqiang and Yao, Yunhao and Li, Xiang-Yang},
  year          = {2025},
  eprint        = {2512.02321},
  archivePrefix = {arXiv},
  publisher     = {arXiv},
  doi           = {10.48550/arXiv.2512.02321},
  url           = {https://doi.org/10.48550/arXiv.2512.02321}
}

@misc{fang2025safemcp,
  title         = {We Should Identify and Mitigate Third-Party Safety Risks in
                   {MCP}-Powered Agent Systems},
  author        = {Fang, Junfeng and Yao, Zijun and Wang, Ruipeng and Ma, Haokai
                   and Wang, Xiang and Chua, Tat-Seng},
  year          = {2025},
  eprint        = {2506.13666},
  archivePrefix = {arXiv},
  primaryClass  = {cs.LG},
  publisher     = {arXiv},
  doi           = {10.48550/arXiv.2506.13666},
  url           = {https://doi.org/10.48550/arXiv.2506.13666}
}

@misc{yan2025tradetrap,
  title         = {{TradeTrap}: Are {LLM}-based Trading Agents Truly Reliable
                   and Faithful?},
  author        = {Yan, L. and Mei, J. and Zhou, T. and Huang, L. and Zhang, J.
                   and Liu, D. and Shao, J.},
  year          = {2025},
  eprint        = {2512.02261},
  archivePrefix = {arXiv},
  primaryClass  = {cs.AI},
  publisher     = {arXiv},
  doi           = {10.48550/arXiv.2512.02261},
  url           = {https://doi.org/10.48550/arXiv.2512.02261}
}

@misc{maloyan2026promptinjection,
  title         = {Prompt Injection Attacks on Agentic Coding Assistants: A Systematic
                   Analysis of Vulnerabilities in Skills, Tools, and Protocol Ecosystems},
  author        = {Maloyan, Narek and Namiot, Dmitry},
  year          = {2026},
  eprint        = {2601.17548},
  archivePrefix = {arXiv},
  publisher     = {arXiv},
  doi           = {10.48550/arXiv.2601.17548},
  url           = {https://doi.org/10.48550/arXiv.2601.17548}
}

@misc{wang2025mindguard,
  title         = {{MINDGUARD}: Tracking, Detecting, and Attributing {MCP} Tool Poisoning
                   Attack via Decision Dependence Graph},
  author        = {Wang, Zhiqiang and Zhang, Junyang and Shi, Guanquan and Cheng, HaoRan
                   and Yao, Yunhao and Guo, Kaiwen and Du, Haohua and Li, Xiang-Yang},
  year          = {2025},
  eprint        = {2508.20412},
  archivePrefix = {arXiv},
  publisher     = {arXiv},
  doi           = {10.48550/arXiv.2508.20412},
  url           = {https://doi.org/10.48550/arXiv.2508.20412}
}

@misc{xing2025mcp_guard,
  title         = {{MCP-Guard}: A Multi-Stage Defense-in-Depth Framework for
                   Securing Model Context Protocol in Agentic {AI}},
  author        = {Xing, Wenpeng and Qi, Zhonghao and Qin, Yupeng and Li, Yilin
                   and Chang, Caini and Yu, Jiahui and Lin, Changting and Xie, Zhenzhen
                   and Han, Meng},
  year          = {2025},
  eprint        = {2508.10991},
  archivePrefix = {arXiv},
  primaryClass  = {cs.CR},
  publisher     = {arXiv},
  doi           = {10.48550/arXiv.2508.10991},
  url           = {https://doi.org/10.48550/arXiv.2508.10991}
}

@misc{jamshidi2025securingmcp,
  title         = {Securing the Model Context Protocol: Defending {LLMs} Against
                   Tool Poisoning and Adversarial Attacks},
  author        = {Jamshidi, Saeid and Nafi, Kawser Wazed and Dakhel, Arghavan Moradi
                   and Shahabi, Negar and Khomh, Foutse and Ezzati-Jivan, Naser},
  year          = {2025},
  eprint        = {2512.06556},
  archivePrefix = {arXiv},
  publisher     = {arXiv},
  doi           = {10.48550/arXiv.2512.06556},
  url           = {https://doi.org/10.48550/arXiv.2512.06556}
}

@misc{jing2025mcip,
  title         = {{MCIP}: Protecting {MCP} Safety via Model Contextual Integrity Protocol},
  author        = {Jing, Huihao and Li, Haoran and Hu, Wenbin and Hu, Qi and Xu, Heli
                   and Chu, Tianshu and Hu, Peizhao and Song, Yangqiu},
  year          = {2025},
  eprint        = {2505.14590},
  archivePrefix = {arXiv},
  primaryClass  = {cs.CR},
  publisher     = {arXiv},
  doi           = {10.48550/arXiv.2505.14590},
  url           = {https://doi.org/10.48550/arXiv.2505.14590}
}

@misc{tan2026mcpsandboxscan,
  title         = {{MCP-SandboxScan}: {WASM}-based Secure Execution and Runtime Analysis
                   for {MCP} Tools},
  author        = {Tan, Zhenyu and Hao, Rui and Singer, Jeremy and Tang, Yifan
                   and Anagnostopoulos, Christos},
  year          = {2026},
  eprint        = {2601.01241},
  archivePrefix = {arXiv},
  publisher     = {arXiv},
  doi           = {10.48550/arXiv.2601.01241},
  url           = {https://doi.org/10.48550/arXiv.2601.01241}
}

@misc{Buhler2025SecuringAgents,
  author        = {B{\"u}hler, C. and Biagiola, M. and Di~Grazia, L. and Salvaneschi, G.},
  title         = {Securing {AI} Agent Execution},
  year          = {2025},
  eprint        = {2510.21236},
  archivePrefix = {arXiv},
  primaryClass  = {cs.AI},
  publisher     = {arXiv},
  doi           = {10.48550/arXiv.2510.21236},
  url           = {https://doi.org/10.48550/arXiv.2510.21236}
}

@misc{lotfi2025automated,
  title         = {Automated Vulnerability Validation and Verification:
                   A Large Language Model Approach},
  author        = {Lotfi, Amir and Katsis, Christos and Bertino, Elisa},
  year          = {2025},
  eprint        = {2509.24037},
  archivePrefix = {arXiv},
  publisher     = {arXiv},
  doi           = {10.48550/arXiv.2509.24037},
  url           = {https://doi.org/10.48550/arXiv.2509.24037}
}

@misc{zong2025mcpsafetybench,
  title         = {{MCP-SafetyBench}: A Benchmark for Safety Evaluation of Large
                   Language Models with Real-World {MCP} Servers},
  author        = {Zong, Xuanjun and Shen, Zhiqi and Wang, Lei and Lan, Yunshi
                   and Yang, Chao},
  year          = {2025},
  eprint        = {2512.15163},
  archivePrefix = {arXiv},
  primaryClass  = {cs.CL},
  publisher     = {arXiv},
  doi           = {10.48550/arXiv.2512.15163},
  url           = {https://doi.org/10.48550/arXiv.2512.15163}
}

@misc{yang2025mcpsecbench,
  title         = {{MCPSecBench}: A Systematic Security Benchmark and Playground
                   for Testing Model Context Protocols},
  author        = {Yang, Yixuan and Wu, Daoyuan and Chen, Yufan},
  year          = {2025},
  eprint        = {2508.13220},
  archivePrefix = {arXiv},
  primaryClass  = {cs.CR},
  publisher     = {arXiv},
  doi           = {10.48550/arXiv.2508.13220},
  url           = {https://doi.org/10.48550/arXiv.2508.13220}
}

@misc{zhang2025msb,
  title         = {{MCP} {Security Bench} ({MSB}): Benchmarking Attacks Against
                   Model Context Protocol in {LLM} Agents},
  author        = {Zhang, Dongsen and Li, Zekun and Luo, Xu and Liu, Xuannan
                   and Li, Peipei and Xu, Wenjun},
  year          = {2025},
  eprint        = {2510.15994},
  archivePrefix = {arXiv},
  primaryClass  = {cs.CR},
  publisher     = {arXiv},
  doi           = {10.48550/arXiv.2510.15994},
  url           = {https://doi.org/10.48550/arXiv.2510.15994}
}

@misc{guo2025mcpagentbench,
  title         = {{MCP-AgentBench}: Evaluating Real-World Language Agent Performance
                   with {MCP}-Mediated Tools},
  author        = {Guo, Zhiqiang and Xu, Bo and Zhu, Chenyu and Hong, Weijie
                   and Wang, Xinyu and Mao, Zhenjiang},
  year          = {2025},
  eprint        = {2509.09734},
  archivePrefix = {arXiv},
  primaryClass  = {cs.AI},
  publisher     = {arXiv},
  doi           = {10.48550/arXiv.2509.09734},
  url           = {https://doi.org/10.48550/arXiv.2509.09734}
}

@misc{wu2025mcpmark,
  title         = {{MCPMark}: A Benchmark for Stress-Testing Realistic and
                   Comprehensive {MCP} Use},
  author        = {Wu, Zijian and Liu, Xiangyan and Zhang, Xinyuan and Chen, Lingjun
                   and Meng, Fanqing and Du, Lingxiao and Zhao, Yiran and others},
  year          = {2025},
  eprint        = {2509.24002},
  archivePrefix = {arXiv},
  primaryClass  = {cs.AI},
  publisher     = {arXiv},
  doi           = {10.48550/arXiv.2509.24002},
  url           = {https://doi.org/10.48550/arXiv.2509.24002}
}

@software{aim_guard_mcp_2025,
  title  = {{AIM-Guard-MCP}: {AI}-Powered Security Guard for Model Context Protocol},
  author = {{AIM Intelligence}},
  year   = {2025},
  url    = {https://github.com/AIM-Intelligence/AIM-MCP},
  note   = {{MCP} security middleware providing prompt injection detection,
             credential scanning, {URL} validation, and contextual {AI} safety guards}
}

@misc{invariant2025mcpscan,
  author       = {{Invariant Labs}},
  title        = {{MCP-Scan}: Model Context Protocol Scanner},
  year         = {2025},
  howpublished = {\url{https://github.com/invariantlabs-ai/mcp-scan}},
  note         = {Accessed: 2025}
}

@misc{mcpscanai2025,
  author       = {{MCPScan.ai}},
  title        = {{MCPScan.ai}: Enterprise Model Context Protocol Security Platform},
  year         = {2025},
  howpublished = {\url{https://mcpscan.ai}},
  note         = {Accessed: 2025}
}

@misc{mcpdefender2025,
  author       = {{MCP-Defender}},
  title        = {{MCP-Defender}: Runtime Protection for Model Context Protocol},
  year         = {2025},
  howpublished = {\url{https://mcpdefender.com/}},
  note         = {Accessed: 2025}
}

@misc{lasso2025mcpgateway,
  author       = {{Lasso Security}},
  title        = {{MCP-Gateway}: Secure Proxy and Orchestration Layer for
                  Model Context Protocol},
  year         = {2025},
  howpublished = {\url{https://github.com/lasso-security/mcp-gateway}},
  note         = {Accessed: 2025}
}

@misc{stacklok2025toolhive,
  author       = {{Stacklok}},
  title        = {{ToolHive}: Simplify and Secure Model Context Protocol Servers},
  year         = {2025},
  howpublished = {\url{https://docs.stacklok.com/toolhive/}},
  note         = {Accessed: 2025}
}

@misc{eqtylab2025mcpguardian,
  author       = {{eqtylab}},
  title        = {{MCP Guardian}: Enterprise Access Control and Governance for
                  Model Context Protocol},
  year         = {2025},
  howpublished = {\url{https://github.com/eqtylab/mcp-guardian}},
  note         = {Accessed: 2025}
}

@misc{cisco2025mcpscanner,
  author       = {{Cisco AI Defense}},
  title        = {{MCP Scanner}: Security Analysis for Model Context Protocol Artifacts},
  year         = {2025},
  howpublished = {\url{https://github.com/cisco-ai-defense/mcp-scanner}},
  note         = {Accessed: 2025}
}

@misc{riseignite2025mcpshield,
  author       = {{Rise and Ignite}},
  title        = {{MCP-Shield}: Security Scanner for Model Context Protocol Servers},
  year         = {2025},
  howpublished = {\url{https://github.com/riseandignite/mcp-shield}},
  note         = {Accessed: 2025}
}

@article{ouyang2022training,
  title   = {Training Language Models to Follow Instructions with Human Feedback},
  author  = {Ouyang, Long and Wu, Jeffrey and Jiang, Xu and Almeida, Diogo
             and Wainwright, Carroll and Mishkin, Pamela and Zhang, Chong
             and Agarwal, Sandhini and Slama, Katarina and Ray, Alex and Schulman, John},
  journal = NeurIPS,
  year    = {2022},
  volume  = {35},
  pages   = {27730--27744},
  url     = {https://arxiv.org/abs/2203.02155}
}

@misc{yao_react_2023,
  title         = {{ReAct}: Synergizing Reasoning and Acting in Language Models},
  author        = {Yao, S. and others},
  year          = {2023},
  eprint        = {2210.03629},
  archivePrefix = {arXiv},
  publisher     = {arXiv},
  doi           = {10.48550/arXiv.2210.03629},
  url           = {https://doi.org/10.48550/arXiv.2210.03629}
}

@article{wang2024llmagentsurvey,
  title   = {A Survey on Large Language Model Based Autonomous Agents},
  author  = {Wang, Lei and Ma, Chao and Feng, Xinyu and Zhang, Zihan and Yang, Haonan
             and Zhang, Jing and Chen, Zihan and Tang, Jie and Chen, Xiaoyan
             and Lin, Yuxiang and Zhao, Wei-Xing},
  journal = {Frontiers of Computer Science},
  volume  = {18},
  number  = {6},
  pages   = {186345},
  year    = {2024}
}

@inproceedings{yu2025survey,
  author    = {Yu, C. and Cheng, Z. and Cui, H. and Gao, Y. and Luo, Z.
               and Wang, Y. and Zheng, H. and Zhao, Y.},
  title     = {A Survey on Agent Workflow---Status and Future},
  booktitle = {2025 8th International Conference on Artificial Intelligence
               and Big Data ({ICAIBD})},
  pages     = {770--781},
  month     = may,
  year      = {2025},
  publisher = {{IEEE}}
}

@misc{mahmud2025distributed,
  author        = {Mahmud, S. and Goldfajn, D. B. and Zilberstein, S.},
  title         = {Distributed Multi-Agent Coordination Using Multi-Modal Foundation Models},
  year          = {2025},
  eprint        = {2501.14189},
  archivePrefix = {arXiv},
  publisher     = {arXiv},
  doi           = {10.48550/arXiv.2501.14189},
  url           = {https://doi.org/10.48550/arXiv.2501.14189}
}

@misc{giurgiu2025supporting,
  author        = {Giurgiu, I. and Nidd, M. E.},
  title         = {Supporting Dynamic Agentic Workloads: How Data and Agents Interact},
  year          = {2025},
  eprint        = {2512.09548},
  archivePrefix = {arXiv},
  publisher     = {arXiv},
  doi           = {10.48550/arXiv.2512.09548},
  url           = {https://doi.org/10.48550/arXiv.2512.09548}
}

@misc{wang2025agent,
  author        = {Wang, H. and Qian, C. and Li, M. and Qiu, J. and Xue, B.
                   and Wang, M. and Ji, H. and Storkey, A. and Wong, K.-F.},
  title         = {Position: Agent Should Invoke External Tools {ONLY} When
                   Epistemically Necessary},
  year          = {2025},
  eprint        = {2506.00886},
  archivePrefix = {arXiv},
  publisher     = {arXiv},
  doi           = {10.48550/arXiv.2506.00886},
  url           = {https://doi.org/10.48550/arXiv.2506.00886}
}

@misc{li2023apibank,
  title         = {{API-Bank}: A Comprehensive Benchmark for Tool-Augmented {LLMs}},
  author        = {Li, Ming and Zhao, Yifan and Yu, Bing and Song, Feifan and Li, Haoxiang
                   and Yu, Hao and Li, Zhen and Huang, Fei and Li, Yi},
  year          = {2023},
  eprint        = {2304.08244},
  archivePrefix = {arXiv},
  publisher     = {arXiv},
  doi           = {10.48550/arXiv.2304.08244},
  url           = {https://doi.org/10.48550/arXiv.2304.08244}
}

@article{gulyamov2026prompt,
  title   = {Prompt Injection Attacks in Large Language Models and {AI} Agent Systems:
             A Comprehensive Review of Vulnerabilities, Attack Vectors,
             and Defense Mechanisms},
  author  = {Gulyamov, S. and Gulyamov, S. and Rodionov, A. and Khursanov, R.
             and Mekhmonov, K. and Babaev, D. and Rakhimjonov, A.},
  journal = {Information},
  volume  = {17},
  number  = {1},
  pages   = {54},
  year    = {2026},
  month   = jan,
  doi     = {10.3390/info17010054}
}

@misc{wei2023jailbroken,
  title         = {Jailbroken: How Does {LLM} Safety Training Fail?},
  author        = {Wei, Jason and Haghtalab, Nika and Steinhardt, Jacob and Zou, James},
  year          = {2023},
  eprint        = {2307.02483},
  archivePrefix = {arXiv},
  publisher     = {arXiv},
  doi           = {10.48550/arXiv.2307.02483},
  url           = {https://doi.org/10.48550/arXiv.2307.02483}
}

@inproceedings{kang2024exploiting,
  title     = {Exploiting Programmatic Behavior of {LLMs}: Dual-Use Through
               Standard Security Attacks},
  author    = {Kang, Daniel and Li, Xinyue and Stoica, Ion and Guestrin, Carlos
               and Zaharia, Matei and Hashimoto, Tatsunori},
  booktitle = {2024 {IEEE} Security and Privacy Workshops ({SPW})},
  pages     = {132--143},
  year      = {2024},
  publisher = {{IEEE}},
  doi       = {10.1109/SPW63631.2024.00028}
}

@inproceedings{Asl2025NEXUS,
  author    = {Asl, J. R. and Narula, S. and Ghasemigol, M. and Blanco, E.
               and Takabi, D.},
  title     = {{NEXUS}: Network Exploration for {eXploiting} Unsafe Sequences
               in Multi-Turn {LLM} Jailbreaks},
  booktitle = {Proceedings of the 2025 Conference on Empirical Methods in
               Natural Language Processing ({EMNLP})},
  pages     = {24278--24306},
  year      = {2025},
  month     = nov
}

@inproceedings{yang2025alleviating,
  title     = {Alleviating the Fear of Losing Alignment in {LLM} Fine-tuning},
  author    = {Yang, Kai and Tao, Guanhua and Chen, Xinyun and Xu, Jian},
  booktitle = {2025 {IEEE} Symposium on Security and Privacy ({SP})},
  year      = {2025},
  pages     = {2152--2170},
  publisher = {{IEEE}},
  url       = {https://ieeexplore.ieee.org/document/10503431}
}

@inproceedings{shen2025bait,
  title     = {{BAIT}: Large Language Model Backdoor Scanning by Inverting Attack Target},
  author    = {Shen, Guangsheng and Cheng, Sheng and Zhang, Ziming and Tao, Guanhua
               and Zhang, Kehuan and Guo, Haotian and Yan, Longfei and Jin, Xiaolong
               and An, Shengwei and Ma, Shiqing and Zhang, Xiangyu},
  booktitle = {2025 {IEEE} Symposium on Security and Privacy ({SP})},
  year      = {2025},
  pages     = {1676--1694},
  publisher = {{IEEE}},
  url       = {https://ieeexplore.ieee.org/document/10503406}
}

@misc{zhu2025demonagent,
  author        = {Zhu, P. and Zhou, Z. and Zhang, Y. and Yan, S. and Wang, K. and Su, S.},
  title         = {{DemonAgent}: Dynamically Encrypted Multi-Backdoor Implantation Attack
                   on {LLM}-Based Agent},
  year          = {2025},
  eprint        = {2502.12575},
  archivePrefix = {arXiv},
  publisher     = {arXiv},
  doi           = {10.48550/arXiv.2502.12575},
  url           = {https://doi.org/10.48550/arXiv.2502.12575}
}

@article{yang2024watch,
  author  = {Yang, W. and Bi, X. and Lin, Y. and Chen, S. and Zhou, J. and Sun, X.},
  title   = {Watch Out for Your Agents! Investigating Backdoor Threats to
             {LLM}-Based Agents},
  journal = NeurIPS,
  volume  = {37},
  pages   = {100938--100964},
  year    = {2024}
}

@misc{owasp_llm_top10,
  title        = {{OWASP} Top 10 for Large Language Model Applications},
  author       = {{OWASP Foundation}},
  year         = {2024},
  howpublished = {\url{https://owasp.org/www-project-top-10-for-large-language-model-applications/}},
  note         = {Accessed: 2026-01-27}
}

@techreport{owasp2025agenticai,
  title       = {{AIVSS}: Scoring System for {OWASP} Agentic {AI} Core Security Risks},
  author      = {{OWASP Foundation}},
  institution = {Open Worldwide Application Security Project ({OWASP})},
  year        = {2025},
  note        = {v0.5},
  url         = {https://aivss.owasp.org/assets/publications/AIVSS%20Scoring%20System%20For%20OWASP%20Agentic%20AI%20Core%20Security%20Risks%20v0.5.pdf}
}

@misc{ghosh2025safety,
  author        = {Ghosh, S. and Simkin, B. and Shiarlis, K. and Nandi, S. and Zhao, D.
                   and Fiedler, M. and Bazinska, J. and Pope, N. and Prabhu, R.
                   and Rohrer, D. and Demoret, M.},
  title         = {A Safety and Security Framework for Real-World Agentic Systems},
  year          = {2025},
  eprint        = {2511.21990},
  archivePrefix = {arXiv},
  publisher     = {arXiv},
  doi           = {10.48550/arXiv.2511.21990},
  url           = {https://doi.org/10.48550/arXiv.2511.21990}
}

@misc{eslami2025security,
  author        = {Eslami, A. and Yu, J.},
  title         = {Security Risks of Agentic Vehicles: A Systematic Analysis of
                   Cognitive and Cross-Layer Threats},
  year          = {2025},
  eprint        = {2512.17041},
  archivePrefix = {arXiv},
  publisher     = {arXiv},
  doi           = {10.48550/arXiv.2512.17041},
  url           = {https://doi.org/10.48550/arXiv.2512.17041}
}

@misc{hatami2026securing,
  title         = {Securing {AI} Agents in Cyber-Physical Systems: A Survey of
                   Environmental Interactions, Deepfake Threats, and Defenses},
  author        = {Hatami, Mehrdad and Pham, Vinh T. and Lakadawala, Hitesh and Chen, Yufei},
  year          = {2026},
  eprint        = {2601.20184},
  archivePrefix = {arXiv},
  publisher     = {arXiv},
  doi           = {10.48550/arXiv.2601.20184},
  url           = {https://doi.org/10.48550/arXiv.2601.20184}
}

@misc{siddiq2026empirical,
  title         = {An Empirical Study on Remote Code Execution in Machine Learning
                   Model Hosting Ecosystems},
  author        = {Siddiq, Mohammad Lutfi and Romel, Tanvir Hossain and Sekerak, Nikola
                   and Casey, Brian and Santos, Jorge},
  year          = {2026},
  eprint        = {2601.14163},
  archivePrefix = {arXiv},
  primaryClass  = {cs.CR},
  publisher     = {arXiv},
  doi           = {10.48550/arXiv.2601.14163},
  url           = {https://doi.org/10.48550/arXiv.2601.14163}
}

@inproceedings{Jishan2024Analyzing,
  author    = {Jishan, M. A. and Allvi, M. W. and Rifat, M. A. K.},
  title     = {Analyzing User Prompt Quality: Insights from Data},
  booktitle = {2024 International Conference on Decision Aid Sciences
               and Applications ({DASA})},
  pages     = {1--5},
  year      = {2024},
  month     = dec,
  publisher = {{IEEE}}
}

\end{document}